\documentclass[%
aps,
pre,
superscriptaddress,
amsmath,amssymb,
reprint,
showkeys,
]{revtex4-2}

\usepackage{array}
\usepackage{mathtools}
\usepackage[dvipsnames]{xcolor}
\usepackage{lipsum}
\usepackage{tikz}
\usepackage{wrapfig}
\usepackage[T1]{fontenc}
\usepackage[utf8]{inputenc}
\usepackage{amsmath}
\usepackage{makecell}
\usepackage{amssymb}
\usepackage{graphicx}
\usepackage{multirow}
\usepackage{blkarray,booktabs,bigstrut}
\usepackage{placeins}
\usepackage{soul}
\usepackage[
citecolor=blue,
colorlinks,
linkcolor=blue,
urlcolor=blue,
]{hyperref}

\usepackage[normalem]{ulem} 
\usepackage[dvipsnames]{xcolor}
\usepackage{dcolumn}
\usepackage{bm}
\usepackage{tabularx,booktabs}
\usepackage{multirow}
\usepackage{float}
\usepackage{chemist}
\usepackage{chemformula}

\begin{document}
	
	
	\title{{Bayesian eco-evolutionary game dynamics}}
	\author{Arunava Patra}
	\email{arunava20@iitk.ac.in}
	\address{
		Department of Physics,
		Indian Institute of Technology Kanpur,
		Uttar Pradesh 208016, India
	}
	
	\author{Joy Das Bairagya}
	\email{joydas@iitk.ac.in (Corresponding author)}
	\address{
		Department of Physics,
		Indian Institute of Technology Kanpur,
		Uttar Pradesh 208016, India 
	}
	\author{Sagar Chakraborty}
	\email{sagarc@iitk.ac.in}
	\address{
		Department of Physics,
		Indian Institute of Technology Kanpur,
		Uttar Pradesh 208016, India
	}
	%
	%
	
	\begin{abstract}
The symbiotic relationship between the frameworks of classical game theory and evolutionary game theory is well-established. However, evolutionary game theorists have mostly tapped into the classical game of complete information where players are completely informed of all other players' payoffs. Of late, there is a surge of interest in eco-evolutionary interactions where the environment's state is changed by the players' actions which, in turn, are influenced by the changing environment. However, in real life, the information about the true environmental state must pass through some noisy channel (like usually imperfect sensory apparatus of the players) before it is perceived by the players: The players naturally are prone to sometimes perceive the true state erroneously. Given the uncertain perceived environment, the players may adopt bet-hedging kind of strategies in which they play different actions in different perceptions. In a population of such ill-informed players, a player would be confused about the information state of her opponent, and an incomplete information situation akin to a Bayesian game surfaces. In short, we contemplate possibility of natural emergence of symbiotic relationship between the frameworks of Bayesian games and eco-evolutionary games when the players are equipped with inefficient sensory apparatus. Herein, we illustrate this connection using a setup of infinitely large, well-mixed population of players equipped with two actions for exploiting a resource (the environment) at two different rates so that the resource state evolves accordingly. The state of the resource impacts every player's decision of playing particular action. We investigate  continuous state environment in the presence of a Gaussian noisy channel. Employing the formalism of deterministic replicator dynamics, we find that noisy information can be effective in preventing resource from going extinct.
	\end{abstract}    
\maketitle

\section{Introduction}
Concealment of information by design or by nature is ubiquitous. Moreover, often gathering complete information is so costly that an agent with apparatus for sensory perception would prefer to take decisions based on incomplete information about the surrounding environment. Naturally, in a population of strategically interacting agents (or players) who interact with environment for their payoffs, the correctness of the perception about the environment plays a crucial role in the long term future of the populations' strategies  and the state of the exploited environment. While this game-environment feedback---when modelled using deterministic eco-evolutionary dynamics---is already known to have interesting consequences in the presence of complete information~\cite{Weitz2016, Tilman2020, Lin2019}, the literature on effect of incomplete information on the dynamics is rather subdued, if not absent. In this context, we recall that a game (modelling the strategic interactions between the players) in the absence of complete information comes under the purview of Bayesian game theory~\cite{Harsanyi1982, Bonanno2018} which has fruitful applications in social sciences~\cite{Li1985} and computer science~\cite{XinyuJin2013, Mohi2009}. We find it interesting that Bayesian games are usually not found in the literature of evolutionary game theory~\cite{Smith1973, MaynardSmith1974, Smith1988}. However, in the context of eco-evolutionary games, we envisage the opportunity of tapping into the rich underpinning of Bayesian games.

Eco-evolutionary game dynamics is simple insightful way of analyzing effectiveness of different ideas of mitigating the tragedy of the commons (ToC)~\cite{Hardin1968, Ostrom1999}---the overexploitation of environment by selfish players---in a mix of players exploiting the environment at different rates; e.g., researchers have explored consequences of punishment, reward~\cite{Mondal2022}, carrying capacity of population~\cite{DasBairagya2021}, generation-wise non-overlapping population~\cite{SohelMondal2024}, finiteness of population and self-renewing resource~\cite{Bairagya2023}, and time delays in making decision~\cite{Roy2023}. Recently, eco-evolutionary game dynamics has been studied in three interconnected species: predator, prey, and parasite~\cite{Roy2024, Ghosh2024}.  Moreover, evolutionary games subject to environmental changes and noise have been comprehensively reviewed for a physics audiences~\cite{Szab2007, Perc2010} in literature. It is important to point out in the context of the present work that noise, either chaotic~\cite{Perc2006a} or stochastic~\cite{Perc2006b}, may be useful in sustaining cooperation which, one hopes, in turn may help preventing ToC in appropriate eco-evolutionary setup.

In the traditional framework of eco-evolutionary game dynamics, one effectively has a replicator equation~\cite{Taylor1978, Schuster1983, Cressman2014} (modelling the evolutionary dynamics of the frequencies of different strategies) coupled with a logistic equation~\cite{May1975} (modelling the  resource under harvest by the players); the coupling is bidirectional: the state of the resource directly affects the game matrix payoffs (parameters in the replicator equation) and the frequency of the strategies directly affects the resource growth rate (parameter in the logistic equation). Needless to say, the resultant nonlinear dynamical coupled equations show rich dynamical features, like, bistability~\cite{Tilman2020}, limit cycles~\cite{ Tilman2020, Gong2022}, heteroclinic cycles~\cite{Weitz2016}, and chaos~\cite{Mondal2022}. 

We note that the aforementioned eco-evolutionary framework overlooks many other aspects of the real world; e.g., among others, the ones we are specifically interested in this paper are: (i) the state of environment perceived by the players may be erroneous~\cite{Rivoire2011, Barfuss2022, Kleshnina2023} and (ii)  the players may have bet-hedging like strategies~\cite{Slatkin1974, Harvey1987} that are contingent on the state of environment. The bet-hedging strategies have been extensively researched upon in the context of evolutionary games. Among some types of such strategies, the most common ones are: conservative bet-hedging strategy~\cite{Gillespie1974, Einum1999} (which is a risk avoiding  strategy employing the single optimal action that ensures survival in the harsher environment), diversified bet-hedging strategy~\cite{Cohen1966} (which leads to employing multiple actions---some {are} optimal in favorable environment and {others are} in harsher one), and adaptive coin flipping bet hedging~\cite{Cooper1982} (which effectively leads to employing an action based on outcome of a coin flipping). Clearly, changing uncertain environment motivates adoption of bet hedgeing strategies.

Practically, environment of a player is almost always uncertain: Knowing the exact state of the environment is unfeasible because the sensory organ of the players may perceive signal or cue about the environmental state erroneously owing to the noisy information channel~\cite{Shannon1948} through which signals or cues pass. Consequently, in a strategic interaction, a player's harvesting rate (her strategy)---if dependent on environmental state---is influenced by her perceived environmental state, her opponent's perceived environmental state and the opponent's harvesting rate. Moreover, the error in perception may either be independent of the state of environment or be a function of the state. It is not surprising that players would use some kind of bet-hedging strategies in such an complex uncertain environment. 

The perception of environmental state of a player can be taken as the \emph{type} of the player. Of course, what type a player is, is a private information of the player. The possible harvesting rates serve as the set of actions of players. Once the possible payoffs of a player (given her type and action profile) and a prior probability distribution over all allowed type profiles are specified, one completes the requirements needed for creating a Bayesian game. Later in this paper we make use of the fact a pure strategy in such a Bayesian game is a bet-hedging like strategy. We remark that a player's knowledge of the prior distribution owes to the implicit simplifying assumption that players know the structure of noisy information channel: The common prior for the given system is the joint probability distribution of different players' perceptions that is fixed by the channel's properties. While there are many textbooks covering this topic, Appendix~\ref{app:bg} succintly elaborates on it tuned to this paper's context.

The most interesting aspect of this paper is that how an uncertain environment's noisy perception automatically gives rise to a Bayesian game scenario---something not quite common in the evolutionary game theory literature to the best of our knowledge. {The situation of incomplete information is created because, owing to potentially erronous perception of environment, no player knows what her opponent's type is.} What we furthermore find interesting in this direction of research is that the ToC may be mitigated or even completely prevented by controlling the probability of erroneous perception. 
    
\section{Framework} \label{sec:framework}
In a simple eco-evolutionary setup there is a well-mixed (unstructured) infinite population of strategically interacting agents (or players) and a consumable resource (to be synonymously called environment). In this section, with a view to bringing forth an eco-evolutionary Bayesian game, we elevate the setup to include a noisy information channel between the players and their environment.
\subsection{The Bayesian game} \label{sec:Bayesian_game}
Specifically, consider a well-mixed population of players with two available actions---harvesting at high rate ($H$) and harvesting at low rate ($L$). Let the true environmental state be perceived as replete with a probability $p_r$, whose expression depends on the precise nature of the information channel. A player with replete perception is $r$-\emph{type} and the one with depleted perception is $d$-\emph{type}. Now, at instant the entire population may be partitioned into two sets of players---one with action $L$ and other with action $H$. Next, we recall~\cite{Weitz2016,Tilman2020} that in the narrative of prevention of ToC in the setting of strategic symmetric one-shot game between two players, the dominant strategy should be $H$ if the environment is perceived as replete and $L$ if the  environment is perceived as depleted.  Thus, mathematically, the payoff matrices of $r$-\emph{type} and $d$-\emph{type} may, respectively, be cast as
 $$
    ~~\sf M_{r}=~
    \begin{blockarray}{ccc}
    &  {L} & {H} &  \\
    \begin{block}{c[cc]}
    {L}~ & R & S\\
    {H}~ & T& P\\ 
    \end{block}
    \end{blockarray}~
    ~\text{and}
    ~\sf M_{d}=~
    \begin{blockarray}{ccc}
    &  {L} & {H} &  \\
    \begin{block}{c[cc]}
    {L}~ & T& P\\
    {H}~ & R& S\\
    \end{block}
    \end{blockarray}~,
    $$
where imposing ordinal conditions---$T>R$ and $P>S$, ensures that required dominant strategy is realized. (For completeness, let us exemplify the matrices by pointing out that in replete perception state, $T$ is the payoff of player employing action $H$ interacting against her opponent playing action $L$; all other elements of the matrices are to be interpreted along similar lines.) In fact, the dominant strategies in ${\sf M_r}$ and ${\sf M_d}$, respectively, are $H$ and $L$. 

Before proceeding further, we clarify the meaning of harvesting and payoffs. Indeed, harvesting at any rate degrades the resource if the resource does not have any intrinsic growth rate. The shared resource considered here has an intrinsic growth rate. Harvesting at a high rate ($H$) means that the harvesting rate is greater than the resource's intrinsic growth rate, causing degradation of the resource. In contrast, harvesting at a low rate ($L$) means that the harvesting rate is lower than the resource's intrinsic growth rate, resulting in the growth of the resource. Next, the players set the preference for harvesting rate profiles according to their perception which can be either replete or depleted; consequently, the payoff matrices ${\sf M_r}$ and ${\sf M_d}$ respectively corresponding to $r$-\emph{type} and $d$-\emph{type} are constructed following the order of preferences in respective perceptions. Note that the payoffs denoted by parameters $R$, $S$, $T$, and $P$ in the payoff matrices do not describe the acquired resource by harvesting. Instead, they represent a player's order of preferences between the action profiles.
      \begin{figure}
   	\centering
   	\includegraphics[scale=0.8]{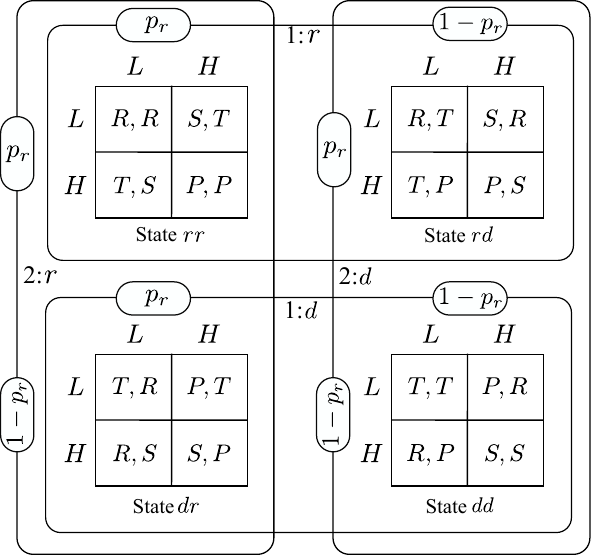}
   	\caption{A two-players--two-type Bayesian game in which each player is uncertain of the other player's type. A frame labeled $i: \alpha$ corresponds to the type $\alpha$ for player $i$; there are four such frames. Also, written in the small boxes over each such frame next to the payoff matrices are the probabilities that type $\alpha$ of player $i$ assigns to each state.}
   	\label{fig:payoff_matrix_Bayesian_game}
   \end{figure}

Since each player can be of two types, there are four different strategic interactions in any \emph{true} state of the environment: ${rr}$, ${rd}$, ${dr}$ and ${dd}$; the second one, e.g., says that the player (say, player-1) is of $r$-type and the opponent (say, player-2) is of $d$-type. We reiterate that a player's type ($r$ or $d$) is determined by her perception (replete or depleted) of the environment---true environmental state does not define a players' type; and the environment, at any point of time, is exactly same for all individuals. Incomplete information about the environment creates an incomplete information scenario about what an opponent perceives as the environmental state: In short, a player does not know the type of her opponent. Consequently, the player does not know whether her opponent's payoff matrix is ${\sf M_r}$ or ${\sf M_d}$; and this uncertainty about the opponent's payoff matrix plays a critical role in formulating the scenario as a Bayesian game as depicted in Fig.~\ref{fig:payoff_matrix_Bayesian_game}.  Obviously, the player-1 with replete perception of the environment finds herself in $rr$ and $rd$ interaction scenarios with probability $p_r$ and $(1-p_r)$, respectively, because she encounters opponent perceiving the same environmental state as replete and deplete with probabilities $p_r$ and $(1-p_r)$, respectively. One can say \emph{equivalently} that \emph{as if} the player-1 assigns \emph{belief} $p_r$ and $(1-p_r)$ on the states $ rr$ and $ rd$. Similarly, one can deduce the other possible beliefs of both the player-1 and the player-2 under different information sets. The reader is referred to Appendix~\ref{app:bg} for more detailed game-theoretic perspective in this context.
 \subsection{The evolutionary Bayesian game dynamics} \label{sec:evol_Bayes_game_dyn}

Given that the environmental state may change over time and with it the perceptions of the players, it is more relevant to partition the population into four sets corresponding to four phenotypes, viz.,  $LL$, $LH$, $HL$, and $HH$. {\color{black}Here, in notation for each phenotype, first letter denotes the action when environmental state is perceived to be depleted and the second letter denotes the action when the state is perceived to be replete.} Three of these phenotypes are bet-hedging like strategies:
\begin{enumerate}
\item{$LL$}: This phenotype harvests at a lower rate irrespective of whether the true environment is perceived as depleted or replete. Thus, it appears as if they choose the strategy conservatively which is optimal for their survival in the true depleted state. This is in line with conservative bet-hedging: They are afraid that their perception might be incorrectly replete when the true environmental state is depleted and  therefore, they adopt $LL$-strategy which is a risk-avoiding strategy that is better suited to the comparatively harsher environment because such a strategy prevents ToC and hence is profitable for the phenotype. Note that the risk involved in a bet-hedging strategy is due to the uncertainty of the true environmental state which, in the case under study, manifests as the uncertainty about the opponent's perception.

\item{$LH$}: This phenotype adopts $L$ action in (perceived) depleted state and $H$ action in (perceived) replete state. This can be interpreted along the line of  adaptive coin-flipping bet hedging strategy. In the presence of temporally evolving environment and its perceiving through an erroneous information channel,  the strategy selection by the player may be thought as if they toss a biased coin---a decision making apparatus they are endowed with---to decide the state of the environment before choosing the corresponding action, $L$ or $H$.
\item{$HL$}: This phenotype adopts $H$ action in (perceived) depleted state and $L$ action in (perceived) replete state.  This has interpretation similar to that for $LH$ strategy.
\end{enumerate}

Let four {strategies} of players---$LL$, $LH$, $HL$, and $HH$---occur with frequencies $ x_{LL}$, $x_{LH}$, $x_{HL}$ and $x_{HH}$, respectively; of course, $ x_{LL}+x_{LH}+x_{HL}+x_{HH}=1$. The payoff matrix, $\sf M_s$, of the normal form symmetric game for these interactions can be easily found using the two-type--two-action Bayesian games (Fig.~\ref{fig:payoff_matrix_Bayesian_game}). Specifically,

\begin{widetext}
		$${\sf M_s}=
		\begin{blockarray}{ccccc}
		&  LL & LH & HL & HH &  \\
		\begin{block}{c[cccc]}
		LL~~ & \makecell{p_rR+(1-p_r)T} & \makecell{p_r\left(p_rR+(1-p_r)S\right)+ \\ (1-p_r) \left(p_rT+(1-p_r)P\right)} & \makecell{p_r\left(p_rS+(1-p_r)R\right) +\\ (1-p_r) \left(p_rP+(1-p_r)T\right)} & \makecell{p_rS+(1-p_r)P} \\	
		 & \makecell{} & \makecell{} & \makecell{} & \makecell{} \\	
		LH~~& \makecell{R} & \makecell{p_rR+(1-p_r)S} & \makecell{p_rS+(1-p_r)R} & \makecell{S} \\	
		& \makecell{} & \makecell{} & \makecell{} & \makecell{} \\	
		HL~~ & \makecell{T} & \makecell{p_rT+(1-p_r)P} & \makecell{p_rP+(1-p_r)T} & \makecell{P} \\	
		& \makecell{} & \makecell{} & \makecell{} & \makecell{} \\	
		HH~~ & \makecell{p_rT+ (1-p_r) R} & \makecell{p_r\left(p_rT+(1-p_r)P\right)+ \\ (1-p_r) \left(p_rR+(1-p_r)S\right)} & \makecell{p_r\left(p_rP+(1-p_r)T\right)+ \\ (1-p_r) \left(p_rS+(1-p_r)R\right)} & \makecell{p_rP+ (1-p_r)S}\\
		\end{block}
		\end{blockarray}\quad.
$$
\end{widetext}
For illustration, let us explain one of the matrix element, say the one corresponding to the strategy profile $(LH,HL)$, i.e., the element at second row and third column. This strategy profile means that the player-1 employs action $L$ and action $H$ on perceiving the environment to be replete and depleted, respectively; while the player-2's actions are just the opposite. {\color{black}Player-1  has a belief that the player-2 considers environment to be replete with probability $p_r$ and depleted with probability $1-p_r$.} Therefore the contribution to the payoff of player-1 is $p_rS+(1-p_r)R$: Note the matrices in the top row of Fig.~\ref{fig:payoff_matrix_Bayesian_game} and the fact that payoff corresponding to action profile $(\star,H)$ is $S$ and that corresponding to action profile $(\star,L)$ is $R$, where $\star$ can be either $L$ or $H$ for the player-1. Since the player-1 herself is $r$-type and $d$-type with probabilities $p_r$ and $1-p_r$, respectively, her net payoff is $p_r[p_rS+(1-p_r)R]+(1-p_r)[p_rS+(1-p_r)R]=p_rS+(1-p_r)R$. Likewise, all other matrix elements are obtained.

Given $\sf M_s$, writing the replicator dynamics governing the evolution so frequencies in the population is straightforward. (For the readers not very conversant with the replicator dynamics, the steps are provided in Appendix~\ref{app:re}.) However, we have found it convenient to work with an alternative equivalent set of variables, $(x,y,D)$, defined through the following coordinate transformation:
\begin{subequations}
\label{eq:newvar}
   \begin{eqnarray}
   	x&\equiv&x_{LL}+x_{LH},
   	\label{eqn:right_type}\\
   	y&\equiv&x_{LL}+x_{HL},
   	\label{eqn: wrong_type}\\
	D&\equiv&x_{LL}-(x_{LL}+x_{LH})(x_{LL}+x_{HL})=x_{LL}-xy.\quad\label{eq:D=}
   \end{eqnarray}
   \end{subequations}
These variables have very useful physical interpretations: $x$ is the frequency of low-harvesters---i.e., the players who adopt action $L$---in (perceived) replete state and $y$ is the frequency of low-harvesters in (perceived) depleted state; and $D$ measures the difference between the frequency of $LL$-strategied players from what the same frequency would be in case frequency of low-harvester in (perceived) depleted state {is} independent of the frequency of the low-harvester in (perceived) replete state.

Finally, the replicator dynamics in terms of the variable given in Eq.~(\ref{eq:newvar}), comes out to be
   \begin{subequations}
   	   \begin{eqnarray}
   	   \dot{x}&=& p_rx(1-x)(f_L-f_H)+(1-p_r) (g_L-g_H)D \label{eqn: evolution_r_type_D_not_0}, \\
   	   \dot{y}&=& (1-p_r)y(1-y)(g_L-g_H)+p_r(f_L-f_H)D  \label{eqn: evolution_d_type_D_not_0},\\
   	   \dot{D}&=&D[p_r(1-2x)(f_L-f_H)+ (1-p_r)(1-2y)(g_L-g_H)] \label{eqn: cov_D}.\nonumber\\
   	   \end{eqnarray}\label{eqn: evolution_D}
   \end{subequations}
Here, $f_L$ and $f_H$, respectively, denote the fitnesses of an $r$-type low-harvester and an {$r$-type} high-harvester; while $g_L$ and $g_H$, respectively, denotes the fitnesses of a $d$-type low-harvester and a $d$-type high-harvester. These are given by
\begin{subequations}\label{eq:fg}
   \begin{eqnarray}
   	f_L&=&g_H= p_r[Rx + S(1-x)]+(1-p_r)[Ry + S(1-y)],\nonumber\\
   	\label{eq:fitness_rL_dH} \\
   	f_H&=&g_L= p_r[Tx + P(1-x)]+(1-p_r)[Ty + P(1-y)].\nonumber\\
   	\label{eq:fitness_rH_dL}
   \end{eqnarray}
   \end{subequations}
     \begin{figure}
   	\centering
   	\includegraphics[scale=1.0]{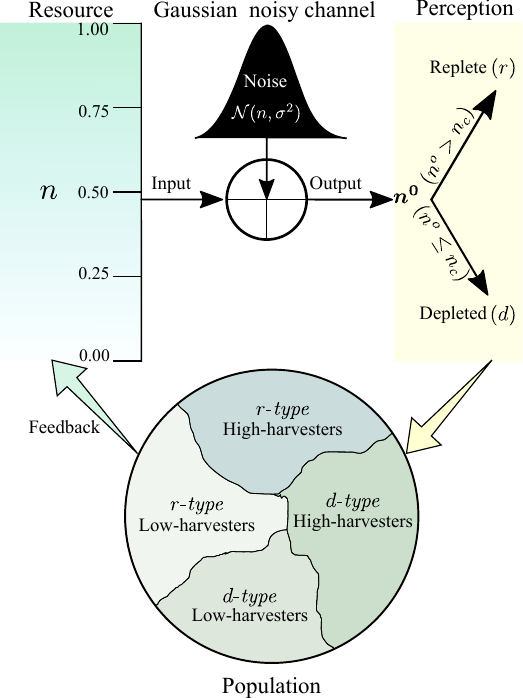}
   	\caption{Schematic diagram illustrating the eco-evolutionary feedback mechanism in the presence of a noisy channel. The continuous resource is illustrated as a colour gradient indicating the resource's abundance state. The signal from the resource passes through a Gaussian noisy channel, where an input $n$ enters the channel and is converted to $n^o$ at the output. The individuals then receive the output $n^o$ and determine the resource's state as either replete or depleted based on their replete perception threshold, $n_c$. Subsequently, individuals engage in either low harvesting $(L)$ or high harvesting $(H)$, resulting in a population consisting of individuals with variety of perceptions and actions. Finally, the existence of feedback of players' actions on the resource is represented by an green arrow.}
	\label{fig:csr}
   \end{figure}
\subsection{The eco-evolutionary Bayesian game dynamics}
Finally, we turn to the crucial aspect of the system---the evolution of the environmental state which has remained implicit in our discussion.
In the game-environment feedback systems, harvesting rates of the individuals {affect} the resource, which in turn affects the net harvesting rate. The exact nature of the environmental states and how they are perceived influence two ingredients of the system: (i) parameter $p_r$ of the information channel and (ii) the evolution equation of the environmental state.

We note that in the setup discussed till now, we have allowed only two types of perception by the individuals: $r$-type and $d$-type. However, in general an environment can have continuous range of states ($n$, say, normalized to range from 0 to 1)---starting from completely depleted state ($n=0$) to completely replete state ($n=1$)---and hence there can have uncountably infinite number of true environmental states. Given a true state $n=n'\in[0,1]$, the incorrectly perceived state by a player can be any other $n\ne n'$, and hence, the possible types of player will be uncountably infinite as well. This is mathematically much harder to handle: not only will be replicator equations become more complicated integro-differential equations~\cite{Cressman2014}, but also the setup of Bayesian games as in Fig.~\ref{fig:payoff_matrix_Bayesian_game} fails as now finite number of payoff matrices won't be enough. For the sake of this paper's goal---which is to bring forth the natural connection between Bayesian games and eco-evolutionary dynamics with incomplete information---it is practical to confine ourselves to a scenario where although the environment remains realistically continuous, the harvesters are allowed to have only two kinds of perception---replete ($n=1$) and depleted ($n=0$). This is achieved by introducing a `replete perception threshold', $n_c$:  A player interprets the true environmental state $n$ as replete when it exceed $n_c$, and as deplete otherwise. We remark that the replete perception threshold does not transform a continuous resource into an equivalent binary resource. Instead, players use this threshold only to interpret the output of the noisy channel (i.e., the perceived abundance) as either replete or depleted, while the resource remains continuous with its precise value being decided by $n$ which---through $p_r$---fixes expected payoff or fitness. A given payoff matrix, however, remains constant across all environmental states as it merely reflect the order of preference corresponding to perception states of the interacting players.

The continuous state environment perception is illustrated through the schematic diagram, Fig.~\ref{fig:csr}. The basic idea is as follows. We consider a resource where the abundance changes continuously. We normalize the abundance of the resource by its carrying capacity and consider relative abundance, $n$,which can take values between 0 to 1: $n=0$ and $n=1$, respectively, correspond to true depleted and true replete states. As a standard model of the noisy sensory apparatus of the individuals, we contemplate a Gaussian noisy channel between the resource and the population. Now, when to call an environment with continuous state values replete and depleted is a matter of convention or perception of a player. To this end, we define a parameter, $n_c\in(0,1)$---replete perception level--- such that an individual recognizes the resource as a replete resource only if the perceived abundance surpasses $n_c$; otherwise the resource is deemed depleted.   

Now if the true state is $n$ then due to incomplete information rendered by noisy channel, a player can perceive the input $n$ as an output, say $n^o$, that has a (truncated) Gaussian distribution with mean $n$ and {a standard deviation, $\sigma$, which characterizes how probable are the larger deviations of output, $n^o$, from the true state.} Thus, the higher the $\sigma$, the more the chances of higher incorrect perception. For a given true state $n$, the probability that the output $n^o$ of the Gaussian noisy channel is less than $n_c$, (i.e., the probability of a player having depleted perception) is proportional to $\int_{0}^{n_c}e^{-\frac{(n^o-n)^2}{2\sigma^2}}dn^o$; the proportionality constant is found by normalizing it.  Thus, in the light of replete perception level, the probability, $\omega(n)$, that an individual is of $d$-type given the exact abundance is $n$ is
   \begin{equation}\label{eq:omegan}
   \omega(n)= \frac{\int_{0}^{n_c}e^{-\frac{(n^o-n)^2}{2\sigma^2}}dn^o}{\int_{0}^{1}e^{-\frac{(n^o-n)^2}{2\sigma^2}}dn^o}=\frac{\rm{erf}\left(\frac{n}{\sigma\sqrt2}\right)+\rm{erf}\left(\frac{n_c-n}{\sigma\sqrt2}\right)}{\rm{erf}\left(\frac{n}{\sigma\sqrt2}\right)+\rm{erf}\left(\frac{1-n}{\sigma\sqrt 2}\right)}.
   \end{equation}%

For the continuously changing resource, it is customary to theoretically work with logistic growth---only change we incorporate here is that the growth rate, $\gamma(x,y,n)$, should depend on the frequencies of low-harvesters of both $r$-type and $d$-type. We, thus, envisage an equation of the form:
\begin{equation}
\dot{n}=\gamma(x,y,n)[n(1-n)].
\label{eqn:resource}
\end{equation}
Note that the frequency of  low-harvesters in the population is given by $(1-\omega(n))x+\omega(n)y$; the frequency of high-harvesters naturally is $1-[(1-\omega(n))x+\omega(n)y]$. Therefore, growth rate of the resource being sustained by low-harvesters and depleted by high-harvesters can be written as 
\begin{equation}
\gamma=\theta_l[(1-\omega)x+\omega y]-\theta_h[1-(1-\omega)x-\omega y],\label{eq:gamma}
\end{equation}
using the proportionality positive constants, $\theta_l$ and $\theta_h$. Therefore, defining $\theta\equiv\theta_l/\theta_h$ and introducing $\epsilon$ to tune the timescale of environmental state's evolution with respect to the replicator dynamics of harvesters, we substitute Eq.~(\ref{eq:gamma}) into Eq.~(\ref{eqn:resource}) to arrive at
\begin{equation}
\dot{n}=\epsilon n(1-n)[-1+(1+\theta)\{(1-\omega)x+\omega y\}].\label{eq:n}
\end{equation}   
   
As the resource dynamics acts back on the evolutionary Bayesian game dynamics~(Eq.~\ref{eqn: evolution_D}), in the case under consideration, $p_r$ should be replaced by $1-\omega(n)$ to ultimately arrive at   
\begin{subequations}
\begin{eqnarray}
\dot{x}&=&[(1-\omega)x(1-x)-\omega D] k(x,y,n),
\label{eq:Resource_contstate_replete_lowharvester}\\
\dot{y} &=&[-\omega y(1-y)+(1-\omega)D]k(x,y,n),
\label{eq:Resource_contstate_deplete_lowharvester}\\
\dot{D}&=&[(1-\omega)(1-2x)-\omega(1-2y)]k(x,y,n)D,\quad \label{eq:D_continuous}
\end{eqnarray}\label{eq: Resource_cont_state}
\end{subequations}
where $k(x,y,n)\equiv(R-S-T+P)[(1-\omega(n))x+\omega(n)y]+(S-P)$. Eq.~(\ref{eq:n}) and Eq.~(\ref{eq: Resource_cont_state}) together describe the eco-evolutionary Bayesian game dynamics in the continuous state environment. We note that while the entries in the payoff matrices of the Bayesian game that describe the payoffs (see Fig.~\ref{fig:payoff_matrix_Bayesian_game}) are independent of the environment even when $p_r$ is replaced by $1-\omega(n)$, the expected payoffs or the fitnesses (see Eq.~\ref{eq:fitness_rL_dH} and Eq.~\ref{eq:fitness_rH_dL}) depend on the true environmental state through $\omega(n)$.

\section{Analyzing eco-evolutionary Bayesian game dynamics}

We perform the linear stability analysis of Eqs.~(\ref{eq:n}) and (\ref{eq: Resource_cont_state}), and tabulate the fixed points $(x^*, y^*, D^*, n^*)$, their existence, and nature of stability in Table~\ref{table: fixed_points}. There are three possible stable fixed points: $(0,1,0,0)$, $(0,1,0,n^{*}_{01})$, and $(0,1,0,1)$. The first two fixed points, viz., $(0,1,0,0)$ and $(0,1,0,n^{*}_{01})$, corresponds to two different kinds of ToC~\cite{Rankin2007}---collapsing ToC and component ToC---respectively. It is obvious that the collapsing ToC corresponds to the complete exhaustion of the resource and if it is partially prevented, then the resulting situation corresponds to the component ToC, where the resource is sustained at a finite state less than its maximum possible state. When the maximum possible state is maintained, the ToC may be said to have been fully averted or prevented: The fixed point $(0,1,0,1)$ corresponds to this scenario. In this paper, when there is no reason for ambiguity, ToC refers to collapsing ToC.  

\begin{table}	
	\caption{The fixed points, $(x^*, y^*, D^*, n^*)$, of the eco-evolutionary dynamics for continuous state environment; and the necessary conditions for the fixed points to exist and to be stable. Here $n^*_{01}$ and $n^*_{10}$ satisfy $\omega(n_{01}^{*})=\frac{1}{1+\theta}$ and $\omega(n_{10}^{*})=\frac{\theta}{1+\theta}$, respectively. Also, $\Omega_0\equiv \frac{1-\omega(0)}{\omega(0)}$ and $\Omega_1\equiv\frac{\omega(1)}{1-\omega(1)}$. }
	\label{table: fixed_points}
	\renewcommand{\arraystretch}{1.6} 
	
	\begin{tabular}{c|c|c}
		\hline 
		\specialrule{0.5pt}{0pt}{0pt}
		\textbf{Fixed point} & \textbf{Existence} & \textbf{Stability} \\
		\hline
		\hline
		
		$(0,0,0,0)$ & Always & Unstable \\
		\hline
		$(0,0,0,1)$ & Always & Unstable \\
		\hline
		\makecell{$(0,1,0,0)$ \\ Collapsing ToC} & Always & Stable if $\Omega_0>\theta$\\
		\hline
		\makecell{$(0,1,0,1)$\\ Prevention of ToC} & Always & Stable if $\Omega_1>{\theta}^{-1}$ \\
		\hline
		$(1,0,0,0)$ & Always & Unstable \\
		\hline
		$(1,0,0,1)$ & Always & Unstable \\
		\hline
		$(1,1,0,0)$ & Always & Unstable \\
		\hline
		$(1,1,0,1)$ & Always & Unstable \\
		\hline
		\makecell{$(0,1,0,n_{01}^{*})$\\ Component ToC} &If $\Omega_1<{\theta}$ and $\Omega_0<\theta^{-1}$ &  Stable   \\
		\hline
		$(1,0,0,n_{10}^{*})$ & If $\Omega_1<{\theta}$ and $\Omega_0<\theta^{-1}$ & Unstable \\
		\hline
		\specialrule{0.5pt}{0pt}{0pt}
	\end{tabular}	
\end{table}

\begin{figure}[h!]
	\centering
	\includegraphics[scale=0.48]{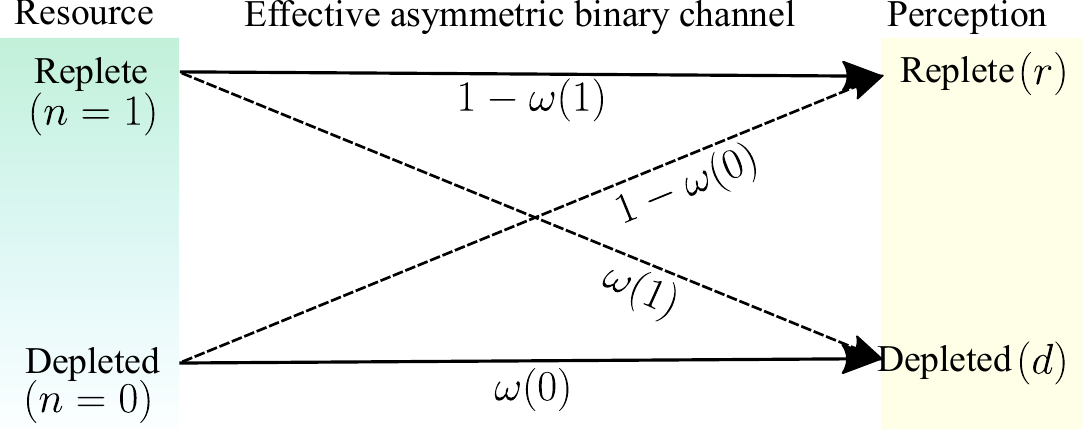}
	\caption{A representation of a continuous resource with a replete perception threshold as a binary environment with an effective asymmetric binary channel. The errors in the channel are $1-\omega(0)$ and $\omega(1)$ in true replete and true depleted environment, respectively. The dashed arrow represents the error in information transfer.}
	\label{fig:cont_assm_channel}
\end{figure}

\begin{figure*}[]
	\centering
	\includegraphics[scale=0.94]{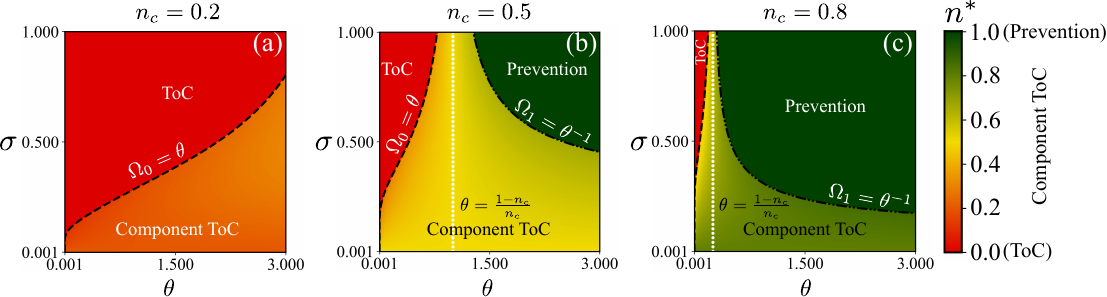}
	\caption{The fate of the resource at asymptotic limit in the continuous resource system is shown in the parameter space $\sigma-\theta$. Subplots (a), (b) and (c) correspond to different replete perception thresholds: $n_c = 0.2$, $n_c = 0.5$ and  $n_c = 0.8$, respectively. The colours red and green, respectively, illustrate the occurrence and prevention of ToC while the gradient transitioning from reddish to greenish represents component ToC. The equations of the dashed and dash-dotted lines are $\Omega_0 = \theta$ and $\Omega_1 = \theta^{-1}$, respectively. These two lines meet at $\theta = \frac{1-n_c}{n_c}$ and $\sigma \rightarrow \infty$, as pointed by the dotted vertical line.}
	\label{fig:Continous_state}
\end{figure*}

The existence and stability of these three fixed points are determined by the parameters $\omega(0)$, $\omega(1)$, and $\theta$, as observed in Table \ref{table: fixed_points}. The parameters, $\omega(0)$ and $\omega(1)$, are the probabilities of an individual perceiving the resource as depleted when the states of the resource are truly depleted and truly replete, respectively. In other words, $1-\omega(0)$ and $\omega(1)$ represent the probabilities of incorrect perception in the (completely) depleted $(n=0)$ and (completely) replete $(n=1)$ states, respectively. Therefore, the outcomes depend only on the (completely) replete $(n=1)$ and (completely) depleted $(n=0)$ states. As a result, the continuous-state environment under consideration can be viewed as an effective two-state system with an asymmetric binary channel between the population and the environment (see Fig.~\ref{fig:cont_assm_channel}).  
The error probabilities in this asymmetric channel are $1-\omega(0)$ and $\omega(1)$ for the true replete and true depleted states, respectively. These error probabilities are influenced by noise parametrized by $\sigma$ (recall Eq.~\ref{eq:omegan}). Since $1-\omega(0)$ and $\omega(1)$ are monotonically increasing functions of $\sigma$ at a fixed value of $n_c$, the relevant parameters characterizing the channel are $\sigma$ and $n_c$. Another parameter characterzing the eco-evolutionary dynamics is $\theta$. Hence, we present the outcomes in Fig.~\ref{fig:Continous_state} in the $\sigma$--$\theta$  space for different values of $n_c$ and try to comprehend them mathematically and physically.

Now in order to better understand the outcomes---collapsing ToC, component ToC, and prevention of ToC---and their dependence on the parameters of interest, we define the ratios of incorrect to correct perception probabilities as $\Omega_0$ and $\Omega_1$, respectively, in the true depleted and the true replete states; i.e., $\Omega_0$ is given by $\frac{1-\omega(0)}{\omega(0)}$ and $\Omega_1$ is given by $\frac{\omega(1)}{1-\omega(1)}$. We also recall Eq.~(\ref{eq:n}), and noting that for every stable fixed point, the state of the population is ($x=0$, $y=1$, $D=0$), the equation at the asymptotic state of the population becomes 
\begin{eqnarray}
\dot{n} = \epsilon n(1-n)[-1+(1+\theta)\omega(n)]. \label{eq:010n}
\end{eqnarray}
The term inside the third bracket determines the growth or the decay, as the multiplicative term $\epsilon n(1-n)$ is always positive. [Eq.~(\ref{eq:010n}) is analogous to the resource state time-evolution equation in Weitz et al.\cite{Weitz2016}, with the frequency of cooperators therein replaced by the frequency of low-harvesters $\omega(n)$. The frequency of low-harvesters is $\omega(n)$ because of assuming $x=0$ and $y=1$.] Note that $\omega(n)$ is a monotonically decreasing function of $n$. Hence, $\omega(0) < \frac{1}{1+\theta}$ implies $\omega(n) < \frac{1}{1+\theta}$ for all $n \in (0, 1]$, and consequently, $\Omega_0 > \theta$ implies $\Omega_1 < \theta^{-1}$; similarly, $\omega(1) > \frac{1}{1+\theta}$ implies $\omega(n) > \frac{1}{1+\theta}$ for all $n \in [0, 1)$, leading to the realization that $\Omega_1 > \theta^{-1}$ implies $\Omega_0 < \theta$. In summary, there are only three independent cases: (A) $\Omega_0 > \theta$, (B) $\Omega_1 > \theta^{-1}$, and (C) $\Omega_0 < \theta$ and $\Omega_1 < \theta^{-1}$. Now, we are ready to discuss the various regions in Fig.~\ref{fig:Continous_state} keeping in mind these three cases.

\subsection {(Colapsing) ToC ($\Omega_0>\theta$)}  
As is clear from Fig.~\ref{fig:Continous_state}, ToC (see the red region in the figure) is effected in the population asymptotically when individuals with depleted and replete perceptions become low-harvesters and high-harvesters, respectively. The ToC occurs regardless of the initial resource state: Even when the resource starts near a true replete state ($n\rightarrow1$). This is because the growth rate ${-1+(1+\theta)\omega(n)}$ (see Eq.~\ref{eq:010n}) becomes negative since $\Omega_0>\theta$ implies $\omega(n)<\frac{1}{1+\theta}$ for all $n$. Consequently, the resource gradually collapses. This argument goes beyond what could be concluded using the linear stability analysis that predicted a linearly (hence, locally) stable fixed point $(0,1,0,0)$---see Table \ref{table: fixed_points}---when $\Omega_0>\theta$. The ToC is  effectively globally stable scenario because this is the only stable fixed point under the condition and our numerical investigation has not divulged existence of any other kinds of attractor like limit cycle or chaotic attractor.
		
We observe that more noise in the Gaussian noisy channel---characterized by high $\sigma$-value leads to ToC regardless of the initial state of the resource and population, and there exists a critical $\theta$ (=$\frac{1-n_c}{n_c}$) below which it appears when $\theta <\Omega_0$, see Fig.~\ref{fig:Continous_state}. (How the critical $\theta$ is found will be discussed in due course.) Since $\Omega_0$ and $\Omega_1$ are monotonically increasing functions of $\sigma$, and describe the ratios of incorrect perception in true depleted and true replete resource state, higher noise increases the proportion of individuals with incorrect perceptions---replete perceptions in a depleted resource and depleted perceptions in a replete resource. Consequently, there are relatively many high-harvesters in a nearly depleted resource, further depleting the resource. On the other hand, even though the frequency of low-harvesters increases with noise in an almost replete resource because the depleted perception individuals increase with noise in true replete resource state, they are insufficient to outcompete the degradation caused by high-harvesters if $\theta$ is low. For this reason, the ToC occurs at lower values of $\theta$, see Fig.~\ref{fig:Continous_state}. Furthermore, an comparative inspection of Figs. \ref{fig:Continous_state}a--c show that the ToC (red) region shrinks as the replete perception level ($n_c$) increases. After all, higher perception increases individuals with depleted perceptions (low-harvesters), which prevents further depletion, making ToC less likely.

\subsection{Prevention of ToC ($\Omega_1>\theta^{-1}$)} 
The understanding of complete prevention of ToC is along the similar lines. Thus, let us focus on the green region (in Fig~\ref{fig:Continous_state}) that indicate fixed point, $(0,1,0,1)$, being stable.  It is interesting that like in the scenario of ToC, in this case also the equilibrium population configuration exclusively consists of low-harvesters with depleted perception and high-harvesters with replete perception. It is the details of the channel that causes a different outcome for the environment. Specifically, we find that complete prevention of ToC is a stable scenario if $\Omega_1>\theta^{-1}$. In fact, the prevention of ToC is effected for almost all possible initial resource states. This is because the growth rate ${-1+(1+\theta)\omega(n)}$ (see Eq.~\ref{eq:010n}) becomes positive as $\Omega_1>\theta^{-1}$ implies $\omega(n)>\frac{1}{1+\theta}$ for all $n$. Ergo, the resource grows unchecked till the resource achieves its maximum possible value, $n=1$.

We observe that high noise (high values of $\sigma$) prevents the ToC when $\theta$ satisfies $\theta >\Omega_1^{-1}$ regardless of the initial state of the resource and population in contrary to the earlier outcomes ToC. High noise increases the individuals with incorrect perceptions in the population because $\Omega_0$ and $\Omega_1$ are monotonically increasing functions of $\sigma$. When the resource is almost replete ($n\rightarrow1$), the increased noise gives rise to more individuals having depleted perception, thereby, more low-harvesters. This surge in low-harvesters becomes advantageous for resources to grow. In contrast, when the resource is nearly true depleted state ($n\rightarrow0$), high noise increases the high-harvesters as replete perception individual increases with it. Then, the growth rate by the low-harvesters must be high enough to outcompete the degradation by high-harvesters. It implies that $\theta$ has to be large enough to realize prevention of ToC, which is what exactly can be seen in Fig.~\ref{fig:Continous_state}. The region of complete prevention enlarges as the replete perception level ($n_c$) rises, see Figs.~\ref{fig:Continous_state}a--c because increasing perception level increases the amount of low-harvesters. That is why even with relatively less values of $\theta$ (at higher value of $n_c$), increasing number of low-harvesters prevents the ToC.
\subsection{Component ToC ($\Omega_0<\theta$ and $\Omega_1<\theta^{-1}$)}

\begin{figure*}
	\centering
	\includegraphics[scale=1.0]{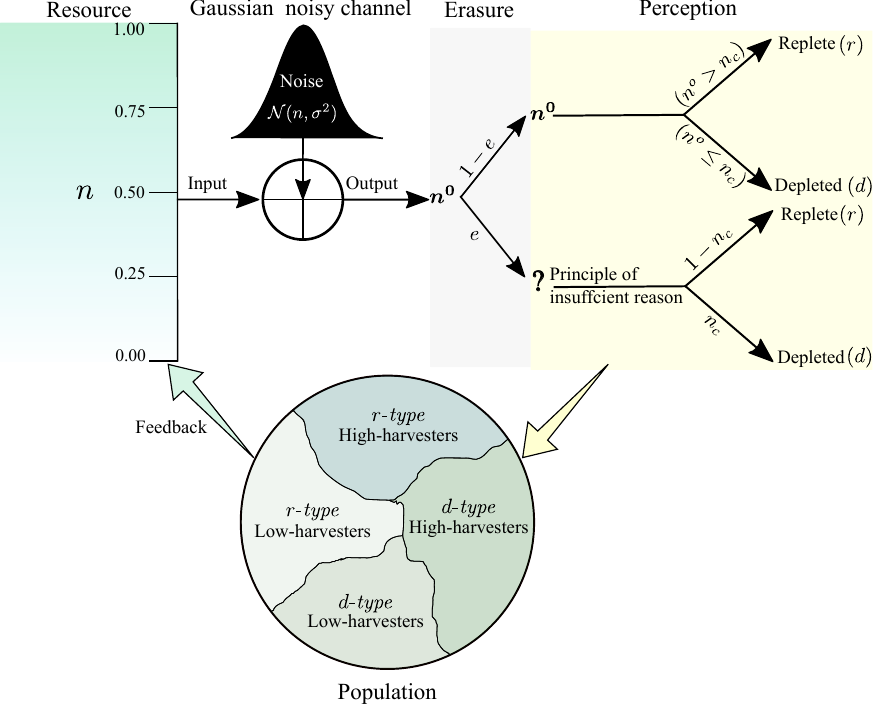}
	\caption{Schematic diagram illustrating a Gaussian noisy channel with erasure. The information is shown to be erased with probability $e$, and erased information is depicted by a question mark. The output, $n^o$, of the Gaussian noisy channel is passed through the erasure with probability $1-e$. Afterwards, an individual perceives the resource based on their replete perception threshold, $n_c$. For the information erased, the individual follows the principle of insufficient reason: They perceive the resource state as replete with probability $1-n_c$ and depleted with probability $n_c$.}
	\label{fig:gauss_erasure_channel}
\end{figure*}
The final case corresponds to the fixed point $(0,1,0,n_{01}^{*})$ which is stable for $\Omega_1<\theta^{-1}$ and $\Omega_0<\theta$. The value of $n_{01}^{*}$ lies between 0 and 1. It means that the resource is neither fully exhausted nor fully abundant, and it is known as component ToC. In Fig.~\ref{fig:Continous_state}, the corresponding region of this component ToC is shown by a gradient of colors depicting  the numerical solution of $\omega(n_{01}^{*})=\frac{1}{1+\theta}$ for different combinations of $\sigma$ and $\theta$. Since the analytical expression of $n_{01}^{*}$  is hard, so rather than arguing what happens in the neighbourhood of the fixed point physically, it is easier to point out that the condition of stability implies that all other fixed points are unstable. Hence, it is obvious that whether one starts near replete environment (ToC prevented) or depleted environment (ToC effected), the state of the environment is pushed away towards the stable component ToC scenario. The detailed mechanism, which we omit in order to avoid trivial repetition, can be understood along the lines discussed in preceding two subsections.
		
Here we point out an interesting observation. Note that when the  noise is at its maximum, i.e., $\sigma=\infty$, the component ToC disappears, and we have either ToC or complete prevention of ToC: observe how the region of component ToC is squeezed about the vertical dotted line, $\theta=\frac{1-n_c}{n_c}$,  in Fig.~\ref{fig:Continous_state} as $\sigma$ increases. This is because the curves, $\Omega_0=\theta$ and $\Omega_1=\theta^{-1}$, meet each other at $\theta=\frac{1-n_c}{n_c}$: As is clear from the definitions and Eq.~(\ref{eq:omegan}), in the limit $\sigma=\infty$, $\Omega_0$ becomes the inverse of $\Omega_1$, and $\Omega_0$ is $\frac{1-n_c}{n_c}$.  Therefore, the condition for getting the component ToC, $\Omega_1<\theta^{-1}$ and $\Omega_0<\theta$ is not achieved at $\sigma=\infty$. The threshold value of $\theta$, i.e., $\frac{1-n_c}{n_c}$, after which the region of complete prevention is found, is inversely proportional to the replete perception level ($n_c$) because the high perception level increases the low-harvesters in the population; therefore, low values of $\theta$ is enough to prevent the ToC at a higher perception level. 

There is a natural way of realising high values of $\sigma$ in the noisy channel: erasure of information as shown schematically in Fig.~\ref{fig:gauss_erasure_channel}. We ponder over the scenario where the information of output in the Gaussian noisy channel is erased with a probability $e$ and the players are unable to determine the environmental state. If the information is erased with a probability $e$, the state is perceived depleted with probability $\omega(n)(1-e)$ and replete with probability $(1-\omega(n))(1-e)$. What the player does with the erased information is arguable. However, a common solution to such a situation is to assume that a player, in the absence of any information about the environmental state, adheres to the `principle of insufficient reason'~\cite{keynes2013treatise}: She assigns a uniform probability distribution upon the state $n\in[0,1]$ and perceives the resource as depleted with probability $n_c$ and as replete with probability $1-n_c$. This makes the Gaussian noisy channel with erasure effectively a Gaussian noisy channel without erasure, such that the state is perceived depleted with probability $\omega(n)(1-e)+en_c$ and replete with probability $(1-\omega(n))(1-e)+e(1-n_c)$ (which equals $1-[\omega(n)(1-e)+en_c]$ as it should). We note that the former and the latter probabilities go to $n_c$ and $1-n_c$, respectively, as $e$ tends to $1$. Consequently, effective $\Omega_1$---the ratio of incorrect to correct perception probabilities in true replete state---is equal to $\frac{n_c}{1-n_c}$ if $e=1$, similar to the case of $\sigma=\infty$. In conclusion, with increasing information erasure, the component ToC is averted and the collapsing ToC is prevented as well for higher values of $\theta$. 
\subsection{Evolutionary Stability}
It is worth emphasizing yet again that in all the three stable scenarios, listed in Table~\ref{table: fixed_points}, lie on a manifold ($D=0$) and the population is composed exclusively of high-harvesters with replete perception ($x=0$) and low-harvesters with depleted perception ($y=1$). The same equilibrium population configuration lead to different states ($n=0$, $n=n^*_{01}$, and $n=1$) of the environment depending on how much error the noisy information channel introduces. 

On the manifold, $D=0$, the frequency of $LL$-strategied players is same as the frequency of low-harvester in perceived depleted state multiplied with the frequency of the low-harvester in perceived replete state, i.e., the frequency of low-harvester in perceived depleted state is independent of the frequency of the low-harvester in perceived replete state. This reminds one of linkage disequilibrium~\cite{Bodmer1967, Ewens1968, Karlin1970, Feldman1979} found in the context of recombination-selection dynamics~\cite{Lewontin1960, Chakraborty2023} in two-locus two-allele scenario~\cite{Hastings1981, Hastings1985, Nagylaki1989}: $D=0$ is to be recognized as the Wright manifold~\cite{Rice1961, hofbauer1998, shahshahani1979new, akin2013geometry, cressman2003book}---the trajectory which begins on this manifold, remains confined on this manifold. Also, in game-theoretic setting, particularly in extensive-form games~\cite{cressman2003book, vonneumann1944}, the Wright manifold corresponds to the set of mixed strategies where decisions made in one subgame do not alter the strategy in other subgames. Coming back to the focus of this section, on $D=0$, the resulting game can be seen as the one given in Fig.~\ref{fig:payoff_matrix_Bayesian_game}.

The equilibrium state of the population may, in fact, be interpreted to be evolutionary stable along the line of the concept of evolutionary stable strategy (ESS)~\cite{Taylor1978,Smith1973}---the cornerstone of evolutionary game theory. Essentially, ESS in the context of population, is a population'state that is robust against invasion by mutants. Since Bayesian evolutionary games {are} not a commonplace in the literature, we take this opportunity to present extension of ESS in the context of this paper---we call the extension, Bayesian evolutionary stable strategy (BESS), for obvious reason; furthermore, we establish that BESS is locally asymptotically stable fixed point of the corresponding replicator dynamic. 

In order to define BESS for the problem in hand, we introdue the notation: $\mathbf{x}=( x_1, x_2)\equiv(x,1-x)$ and $\mathbf{y}=(y_1, y_2)\equiv(y, 1-y)$; obviously, $(\mathbf{x}, \mathbf{y})$ represents the state of the entire population. Let the payoff matrices for type $r$ player playing against another type $r$ player and type $d$ player be $\sf A$ and $\sf B$, respectively. Also, the payoff matrices for type $d$ player playing against another type $r$ player and type $d$ player be $\sf F$ and $\sf G$, respectively. In our context as in Fig.~\ref{fig:payoff_matrix_Bayesian_game}, ${\sf A}={\sf B}={\sf M_r}$ and ${\sf F}={\sf G}={\sf M_d}$. Therefore, we give the following definition:

A state $(\hat{\mathbf{x}},\hat{\mathbf{y}} )$ is a BESS if for every population state, $({\mathbf{x}},{\mathbf{y}} )$, sufficiently close but not equal to $(\hat{\mathbf{x}},\hat{\mathbf{y}})$,
\begin{eqnarray}
	p_r\hat{\mathbf{x}}\{p_r{\sf A}{\mathbf{x}}+(1-p_r){\sf B}{\mathbf{y}}\}&+& (1-p_r)\hat{\mathbf{y}}\{p_r{\sf F}{\mathbf{x}}+(1-p_r){\sf G}{\mathbf{y}}\}\nonumber\\
	&>&\nonumber\\
	p_r{\mathbf{x}}\{p_r{\sf A}{\mathbf{x}}+(1-p_r){\sf B}{\mathbf{y}}\}&+& (1-p_r){\mathbf{y}}\{p_r{\sf F}{\mathbf{x}}+(1-p_r){\sf G}{\mathbf{y}}\}.\nonumber\\
	\label{Cond: ESS}
\end{eqnarray}%
The definition is motivated from the fact that fitness, i.e., expected payoff, of residents of the population at BESS should be higher than that of invading mutants (who perturb the population's resident state) so as to resist the invasion by latter. We note that $p_r=0,1$ leads to the standard definition~\cite{Hofbauer1979} of ESS in games with complete information, as it should.

One can show (see Appendix~\ref{app:3}), by relating BESS to Bayesian Nash equilibrium (BNE)~\cite{Harsanyi1968part2}, that a BESS is fixed point of the replicator equation corresponding to the Bayesian game dynamics. A recast version of Eqs. (\ref{eqn: evolution_r_type_cov}) and (\ref{eqn: evolution_d_type_cov}) on $D=0$ is as follows:
\begin{subequations}
	\begin{eqnarray}
		\dot{x}_i&=& p_rx_i(\mathbf{e}_i-\mathbf{x})(p_r{\sf A}\mathbf{x}+(1-p_r){\sf B}\mathbf{y}) \label{eqn:evolution_r_matrix_form} \\
		\dot{y}_i&=& (1-p_r)y_i(\mathbf{e}_i-\mathbf{y})(p_r{\sf F}\mathbf{x}+(1-p_r){\sf G}\mathbf{y}),
		\label{eqn:evolution_d_matrix_form}
	\end{eqnarray}\label{eqn: evolution_D_0_matrix_form}
\end{subequations}
where $\mathbf{e}_i$'s are unit vectors: $\mathbf{e}_1=(1,0)$ and $\mathbf{e}_2=(0,1)$. Finally, we now are ready to show that the fixed point corresponding to BESS is an asymptotically stable fixed point in Bayesian game dynamics (Eq.~\ref{eqn: evolution_D_0_matrix_form}). 

To this end, we consider the following function:
\begin{equation}
	V(\mathbf{x},\mathbf{y})=p_r\sum_{i=1}^{2}\hat{x}_i\log\left(\frac{\hat{x}_i}{x_i}\right) +(1-p_r)\sum_{i=1}^{2}\hat{y}_i\log\left(\frac{\hat{y}_i}{y_i}\right).
\end{equation}
The function---being a convex combination of two relative entropies~\cite{Cover2005}---must be positive definite for all allowed values of $(\mathbf{x},\mathbf{y})\ne(\hat{\mathbf{x}},\hat{\mathbf{y}})$. In addition, if $V$ is such that $\dot{V}<0 ~\forall (\mathbf{x}, \mathbf{y})\ne (\hat{\mathbf{x}}, \hat{\mathbf{y}})$ in the neighborhood of $(\hat{\mathbf{x}},\hat{\mathbf{y}})$, then $V$ qualifies as a Lyapunov function~\cite{Wiggins2003} which implies that the BESS is locally asymptotically stable fixed point. To see that this indeed is the case, we note that the time derivative of $V$ is given by,
	\begin{eqnarray}
		&&\dot{V}=-\left[ p_r\hat{\mathbf{x}}\{p_r{\sf A}{\mathbf{x}}
		+(1-p_r){\sf B}{\mathbf{y}}\}\right.\nonumber\\
		&&\phantom{\dot{V}=-}+ (1-p_r)\hat{\mathbf{y}}\{p_r{\sf F}{\mathbf{x}}+(1-p_r){\sf G}{\mathbf{y}}\}\nonumber\\
		&&\phantom{\dot{V}=-}-p_r{\mathbf{x}}\{p_r{\sf A}{\mathbf{x}}+(1-p_r){\sf B}{\mathbf{y}}\}\nonumber\\
		&&\phantom{\dot{V}=-}\left.- (1-p_r){\mathbf{y}}\{p_r{\sf F}{\mathbf{x}}+(1-p_r){\sf G}{\mathbf{y}}\}\right],
	\end{eqnarray}
where we have used Eq.~(\ref{eqn: evolution_D_0_matrix_form}). Clearly, $\dot{V}<0 ~\forall (\mathbf{x}, \mathbf{y})\neq (\hat{\mathbf{x}}, \hat{\mathbf{y}})$ in the neighborhood of $(\hat{\mathbf{x}}, \hat{\mathbf{y}})$ when we use inequality (\ref{Cond: ESS}).

We observe that $(x,y)=(0,1)$ is a BESS according to inequality~(\ref{Cond: ESS}) on substituting ${\sf A}={\sf B}={\sf M_r}$, ${\sf F}={\sf G}={\sf M_d}$, and $p_r=1-\omega(n)$. In conclusion, the fact seen in Table~\ref{table: fixed_points}, that all the three stable scenarios lying on $D=0$ renders the population to be in a state such that $x=0$ and $y=1$, is due to this state's being a BESS.

\section{Conclusions}
We have established how game-environment feedback scenario in the presence of noisy perceptions about environment naturally gives rise to Bayesian eco-evolutionary games. The players sometimes have incorrect perception about the environmental state, leading to their incomplete information about others' perceptions. The uncertainty about perception of the environmental states can manifest itself as adoption of bet-hedging like strategies which may be interpreted as pure strategies in the corresponding Bayesian game. (In fact, it gets highlighted that strategic interaction between players with any bet-hedging strategies can, in general, be treated as Bayesian games.) We assume that such strategies are genetically hardwired into the players of the population, leading to the interpretation of the strategies as phenotypes. Consequently, the paradigmatic replicator equation has been used to model the evolution of the frequencies of the phenotypes in the very large (mathematically infinite) population.

Within the simplest non-trivial setup with binary perception states---replete and depleted---we find that the long term dynamics is dependent on relation between $\theta$ and $\Omega_i$ ($i=0,1$), respectively, measuring efficiency of a low-harvester compared to a high-harvester and the ratio of erroneous perception probability to correct perception probability. In the environment with continuous states, the information channel between the players and the environment turns out to be an asymmetric binary channel. Controlling what information about the environment reaches the players mathematically means to modify the channel by tweaking $\Omega_i$, and this can result in mitigating the ToC. In this context, we have shown an interesting result that with increasing information erasure, the component ToC is fully averted and the collapsing ToC is prevented for higher values of $\theta$. 

An important caveat is worth highlighting. In the framework of classical Bayesian game, the players are supposed to be rational and to have conscious subjective beliefs about the state of nature. However, in the Bayesian eco-evolutionary game discussed here, since the frequency distribution of the types of players in the population is determined by the noisy information channel, it can be thought as if beliefs are imparted identically on every player by the channel. Consequently, the framework of Bayesian game has been adapted herein to a population of players who need not have internal conscious beliefs about the other players' perceptions (or types).

In literature there exists a different way to tackle uncertainty about the states of the environment: One may try infer the future environmental states using information from the current perceived state and the players' actions \cite{Tilman2023}. Players thereby form a belief about the future states of the environment, which they update over time. Rational players then estimate their future payoffs based on their current beliefs and make strategic decisions accordingly. In comparison, our framework presented herein involves a rational player making decisions based on the uncertainty about others' preferences.

The framework presented herein can transcend the backstory of mere resource harvesting and the ToC.  For example, one could revisit the setup~\cite{Rand2017} where agents employ automatic process or controlled process such that the cognitive processing alters the environment of the interacting agents, and introduce the fact that the agents have incomplete information about the environment. Also, in society, especially in politics~\cite{Barfar2022}, spreading misinformation to control behaviour of people is very commonplace, and in turn, the behavioural feedback decides the state of the political situation---a typical scenario amenable to Bayesian eco-evolutionary game theoretic modelling. We are hopeful that this paper will initiate further exciting investigations into Bayesian eco-evolutionary games while extending the framework to encompass more realistic case of finite~\cite{Nowak2004} and structured population~\cite{Lin2019, Ohtsuki2006} where one could additionally consider that different individuals gather information through non-identical noisy channels.
\acknowledgements
AP thanks CSIR (India) for the financial support in the form of Senior Research Fellowship. JDB has been supported by Prime Minister's Research fellowship (govt. of India) and SC acknowledges the support from SERB (DST, govt. of India) through project no. MTR/2021/000119. 
\appendix
\section{Noisy channel induced Bayesian game}\label{app:bg}
Here we want to succinctly present how the scenario presented in Sec.~\ref{sec:Bayesian_game} qualifies as a Bayesian game.  

A rather unrealistic and stringent assumption in a game is that of complete information, i.e., strategically interacting rational agents have common knowledge about the basic mathematical structure, viz.,  the utility function and the strategies of each others. In reality, the information available to a player is incomplete: Naturally, the players in such incomplete information game can only have some belief about the payoff relevant data---also called state of the nature. To solve such a game, it is apparent that a player's belief, the belief about opponent's belief, the belief about  opponent's belief about the player's belief, and similar higher order beliefs need to be known. Dealing with such a hierarchy of belief is very inconvenient. 

This problem was circumvented by a pioneering idea of the Nobel laureate Harsanyi~\cite{Harsanyi1967part1, Harsanyi1968part2, Harsanyi1968part3}---the type-centered approach to model the incomplete information game. Under this interpretation, the players can be of different types, and the focal player knows her representing type but does not know the representing type of opponent. However,  players are supposed to know---as a common knowledge---the joint probability distribution of their type and their opponent's type, which is an underlying assumption of the type-centered interpretation, more tersely known as the common prior assumption. Then, the players try to assess the conditional probability of the opponent's types given his/her a particular type by doing probabilistic estimations using the Bayes' updating. The game is termed as a Bayesian game. 

The eco-evolutionary game considered in this paper actually qualifies as a Bayesian game because, owing to the property of the noisy channel being known to every individual in the population, the criterion of existence of a common prior, that is a common knowledge, is satisfied automatically. To understand it more conspicuously, let us now discuss Harsanyi's idea tuned to our purpose.
   
    \begin{figure}[h!]
   	\centering
   	\includegraphics[scale=0.5]{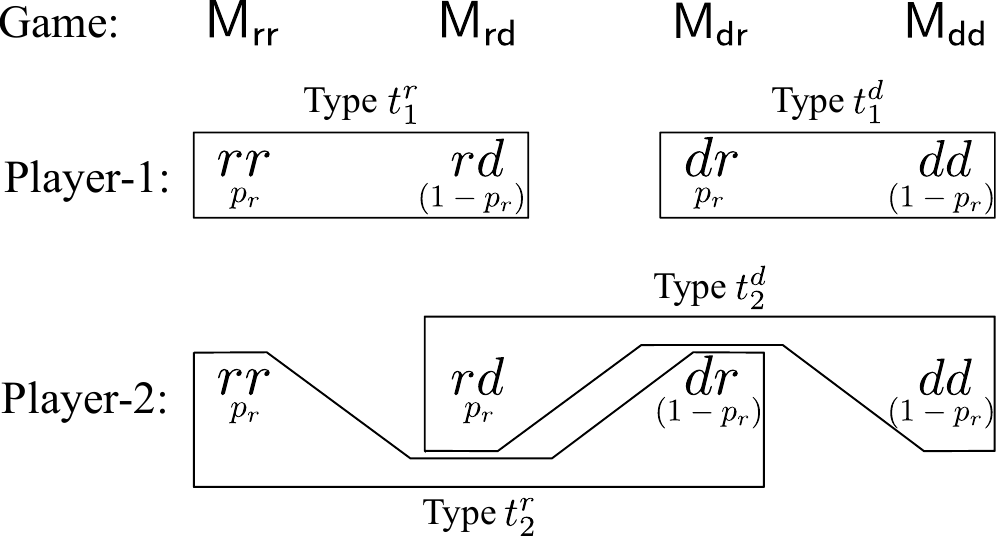}
   	\caption{Schematic diagram shows the transformation from a \emph{state-space} structure to \emph{type-space} structure of a scenario with two-sided incomplete information. There are four states: $rr, rd, dr$, and $dd$. The games $\sf M_{rr}$, $\sf M_{rd}$, $\sf M_{dr}$ and $\sf M_{dd}$ are played in the states $rr, rd, dr$ and $dd$, respectively. For instance, $\sf M_{rd}$ means that the focal player has the payoff matrix, $\sf M_r$, and the opponent has the payoff matrix $\sf M_d$. The player-1 has two information sets---$\{rr, rd\}$ and $\{dr, dd\}$, depicted by boxes. Similarly, the player-2 has two information sets $\{rr, dr\}$ and $\{rd, dd\}$. The players are uncertain about the states in the information set. The beliefs ($p_r$ or $1-p_r$) of players on the states in an information set are written below the states. Each formation set of a player is identified as a type of the player.}
   	\label{fig:Schem_Bayes_game}
   \end{figure}

Now we systematically introduce knowledge-belief structure involved in our model and the transformation from that structure to the type space approach to fit with the Bayesian game. 

In our context, we have four states---$rr$, $ rd$, $ dr$, and $ dd$---as far as binary interaction is concerned. In each state, the first and the second letters correspond to the perception of player-1  and player-2, respectively.  In each state, there corresponds a payoff matrix. Let us assume that the individuals perceive the replete and the depleted state, respectively, with the probabilities, $p_r$ and $(1-p_r)$. Thus, the player-1 has two information sets, $\{ rr, rd\}$ and $\{ dr,dd\}$, while the opponent has the information sets, $\{ rr, dr\}$ and $\{ rd, dd\}$. The players know a particular information set but they are uncertain between the states in the information set. To quantify the uncertainty, each player assigns a probability, i.e., belief, on each of the states inside an information set such that sum of probabilities of these states must be unity; thereby, a knowledge-belief structure is set up for the incomplete information strategic interaction. For example, in an information set $\{ rr, rd\}$ of player-1, she finds the states $ rr$ and $ rd$ with probability $p_r$ and $(1-p_r)$, respectively, because she finds opponent's (player-2) perception replete and depleted with probability $p_r$ and $(1-p_r)$, while her perception remains replete in this information set. (Note that this is so because she knows that all the players have identical information channel to deal with.) One can say equivalently that as if the player-1 assigns the beliefs $p_r$ and $(1-p_r)$ on the states $ rr$ and $ rd$. Similarly, one can calculate the beliefs for the other information set of player-1 and also for the player-2, {\color{black}see Fig.~\ref{fig:Schem_Bayes_game}}. 
   
The idea of transforming from state space structure to a type space approach is the following: Each information set is identified as a type of a player. Therefore, player-1 can be of two types corresponding to the two information sets, $\{ rr, rd\}$ or $\{ dr,dd\}$. We, respectively, label these types by ${ t_1^{ r}}$ and ${ t_1^{  d}}$. Likewise, the player-2 have two types ${ t_2^{ r}}$ and ${ t_2^{ d}}$ corresponding to the information sets $\{ rr, dr\}$ and $\{ rd, dd\}$. As a player knows her own type but not the interacting opponent's type, the uncertainty in a player's mind changes from that of the states to that of the types of opponent;  therefore, the beliefs of the individuals over the states gets converted to the beliefs over opponent's types.
   
The belief of player-1's ${ t_1^{ r}}$ type about the player-2's ${ t_2^{ r}}$ type is $p_r$ because $ rr$ is common state which belongs in the both information sets  $\{ rr, rd\}$ of player-1 and $\{ rr, dr\}$ of player-2: As player-1 assigns the belief $p_r$ on the state $ rr$ in the information set $\{ rr, rd\}$, she finds the information set $\{ rr, dr\}$ of player-2 with probability $p_r$. In similar manner, the beliefs structure over cases for all types of both players can be determined---the belief of type ${ t_1^{ d}}$ on the opponent types ${ t_2^{ r}}$ and  ${ t_2^{ d}}$ is  $p_r$ and $1-p_r$; and the beliefs of both  ${ t_2^{ r}}$ and  ${ t_2^{ d}}$ types of player-2 are $p_r$ and $1-p_r$, respectively, on the types ${ t_1^{ r}}$ and ${ t_1^{ d}}$  of player-1.   
   
Now, most important aspect of the above knowledge-belief structure is that these beliefs are (Harsanyi) consistent beliefs, i.e., there exists a common prior which is a joint probability distribution over all possible types of the players such that the beliefs associated with a given type can be derived from the common prior using the Bayes' updating. Here the common prior, $P\{(t_1^{ r},  t_2^{ r}),(t_1^{ r},  t_2^{ d}),(t_1^{ d},  t_2^{ r}),(t_1^{ d},  t_2^{ d})\}$ = $\{p_r^2, p_r(1-p_r), (1-p_r)p_r, (1-p_r)^2 \}$ from which all types of a player can derive their beliefs about opponent's types. For example, $P(t_1^{ r}| t_2^{ d})=P(t_1^{ r},  t_2^{ r})/P(t_2^d)=p_r(1-p_r)/p_r=1-p_r$; therefore, as mentioned above, it shows that the belief of ${ t_2^{ d}}$ type of player-2 about the type ${ t_1^{ r}}$ of player-1 is $1-p_r$. 

Of course, the underlying assumption of Bayesian game that the common prior is common knowledge to everyone is also satisfied as the property of the unique noisy channel is assumed to be known to everyone.  Thus, the given scenario fulfils the requirement of a Bayesian game. Finally, we point out that in the main text of this paper, the types $t_1^{ r}$  and $t_2^{ r}$ have been conveniently denoted as $r$, the types  $t_1^{ d}$ and $t_2^{ d}$ are denoted by $d$, and $p_r$ is taken to be $1-\omega(n)$. 
\section{Replicator Dynamics}\label{app:re}
Here we provide the steps leading to Eq.~(\ref{eqn: evolution_D}). Having recast the two-type--two-action Bayesian game (displayed in Fig~\ref{fig:payoff_matrix_Bayesian_game}) into a four-strategy normal form game with symmetric payoff matrix, $\sf M_s$, as given in Sec.~\ref{sec:evol_Bayes_game_dyn}, the replicator equation~\cite{Taylor1978}---in line with the Darwinian tenet being followed in an unstructured infinite population where the players randomly interact with each other---is written as follows:
   \begin{eqnarray}\label{eqn: evolution_all}
      	\dot{x}_{ij}&=&x_{ij}(f_{ij}-\bar{f}), 
   \end{eqnarray}%
   with $i,j\in{\{L,H\}}$ and $\sum_{i\in{\{L,H\}}} \sum_{j\in{\{L,H\}}}x_{ij}=1$.
Here we denote the fitness of $ij$-strategied player by $f_{ij}$. These individual fitnesses can be found from the payoff matrix, $\sf M_s$. Specifically, $f_{ij}=\sum_{k\in{\{L,H\}}} \sum_{l\in{\{L,H\}}}x_{kl}m_{ij,kl}$, where $m_{ij,kl}$ denotes that element of $\sf M_s$ which $ij$-th strategied player gets on interacting with a $kl$-th strategied player. Thus, the average fitness, $\bar{f}$, of a player in the population is $\sum_{i\in{\{L,H\}}} \sum_{j\in{\{L,H\}}}x_{ij}f_{ij}$.

 Next, with the change of variable as given in Eq.~(\ref{eq:newvar}), we concentrate on the frequencies---$x$ and $y$---of $L$, when the corresponding types are $r$ and $d$, respectively. It is obvious through Fig~\ref{fig:payoff_matrix_Bayesian_game} that for $r$ type, the fitnesses of $L$ and $H$ actions---respectively, denoted by $f_L$ and $f_H$; and for $d$ type, the fitnesses of $L$ and $H$ actions---respectively, denoted by $g_L$ and $g_H$, are given by Eq.~(\ref{eq:fg}). Consequently, it follows that
   \begin{eqnarray}\label{eq:ffg}
   	f_{ij}&=& p_rf_i+(1-p_r)g_j,
   \end{eqnarray}
 where $i,j\in{\{L,H\}}$. Now, using this equation, and Eqs. (\ref{eqn:right_type}) and (\ref{eqn: wrong_type}), in the definition of $\bar{f}$ written above, we arrive at an equivalent expression of $\bar{f}$:
   \begin{eqnarray}\label{eq:fbar}
   	\bar{f}&=& p_rxf_L+p_r(1-x)f_H+(1-p_r)yg_L\quad\nonumber\\
	\phantom{\bar{f}}&\phantom{=}&+(1-p_r)(1-y)g_H.
   \end{eqnarray}

Subsequently,  to explicitly follow the time evolution of frequency of an action for a given type, we take the time derivative both side of Eqs. (\ref{eqn:right_type}) and (\ref{eqn: wrong_type}), and use Eqs.~(\ref{eqn: evolution_all}),~(\ref{eq:ffg}), and (\ref{eq:fbar}) to further simplify them to arrive at
   \begin{subequations}
   \begin{eqnarray}
   	\dot{x}&=& p_rx(1-x)(f_L-f_H)\nonumber\\ 
   	&&+(1-p_r)[g_L(x_{LL}-xy)+g_H(x_{LH}-x(1-y))],\nonumber\quad\\
   	\label{eqn: evolution_r_type_cov}\\
   	\dot{y}&=& (1-p_r)y(1-y)(g_L-g_H)\nonumber\\ &&+p_r[f_L(x_{LL}-xy)+f_H(x_{HL}-(1-x)y)]. \label{eqn: evolution_d_type_cov}
   \end{eqnarray}\label{eqn: evolution_cov}
   \end{subequations}%
Here the terms $x_{LL}-xy, x_{LH}-x(1-y)$, and {$ x_{HL}-(1-x)y$} are the covariance terms; let us, for brevity, denote them as $\sigma_{LL}, \sigma_{LH}$ and $\sigma_{HL}$, respectively. One can easily show that $\sigma_{LL}=-\sigma_{HL}=-\sigma_{LH}$, implying that a single variable, $D\equiv\sigma_{LL}$, can be used to describe these terms. Consequently, Eq.~(\ref{eq:Resource_contstate_replete_lowharvester}) and Eq.~(\ref{eq:Resource_contstate_deplete_lowharvester}) follow. Finally, we note that Eq.~(\ref{eqn: evolution_all}) has three independent variables; hence, we need one more variable to completely describe the whole system. As per Eq.~(\ref{eq:newvar}), $D$ is one such variable. Using different expressions found till now, the time derivative of Eq.~(\ref{eq:D=}) can be similarly simplified to arrive at Eq.~(\ref{eq:D_continuous}) after some straightforward algebraic steps.
\section{BESS implies BNE and fixed point}\label{app:3}
We remind ourselves that BNE~\cite{Harsanyi1968part2} is a strategy profile (a set containing a strategy for each player type) such that no type has any incentive to deviate from her strategy given her beliefs about the others.

To show that BESS implies BNE, we consider a neighborhood state $(\mathbf{x}, \mathbf{y})$ of BESS $(\hat{\mathbf{x}},\hat{\mathbf{y}})$, where $\mathbf{x}\equiv(1-\epsilon)\hat{\mathbf{x}}+\epsilon\mathbf{x'}$ and $\mathbf{y}\equiv(1-\epsilon)\hat{\mathbf{y}}+\epsilon\mathbf{y'}$ $\forall \epsilon~ \text{and}~ \forall (\mathbf{x'}, \mathbf{y'})\neq(\hat{\mathbf{x}},\hat{\mathbf{y}} ), \text{with}~ 0<\epsilon<\epsilon_{(\mathbf{x}, \mathbf{y})}$ if there exists an invasion barrier $\epsilon_{(\mathbf{x}, \mathbf{y})}>0$. If we put $(\mathbf{x}, \mathbf{y})$ into Eq. (\ref{Cond: ESS}) and the condition transforms as,  

\begin{widetext}
\begin{eqnarray}
	p_r\hat{\mathbf{x}}\{p_r{\sf A}\hat{\mathbf{x}}+(1-p_r){\sf B}\hat{\mathbf{y}}\}+ (1-p_r)\hat{\mathbf{y}}\{p_r{\sf F}\hat{\mathbf{x}}+(1-p_r){\sf G}\hat{\mathbf{y}}\}
	>
	p_r{\mathbf{x'}}\{p_r{\sf A}\hat{\mathbf{x}}+(1-p_r){\sf B}\hat{\mathbf{y}}\}+ (1-p_r){\mathbf{y'}}\{p_r{\sf F}\hat{\mathbf{x}}+(1-p_r){\sf G}\hat{\mathbf{y}}\},\nonumber
\end{eqnarray}%
up to first order in $\epsilon$. However, if 
\begin{eqnarray}
	p_r\hat{\mathbf{x}}\{p_r{\sf A}\hat{\mathbf{x}}+(1-p_r){\sf B}\hat{\mathbf{y}}\}+ (1-p_r)\hat{\mathbf{y}}\{p_r{\sf F}\hat{\mathbf{x}}+(1-p_r){\sf G}\hat{\mathbf{y}}\}
	=
	p_r{\mathbf{x'}}\{p_r{\sf A}\hat{\mathbf{x}}+(1-p_r){\sf B}\hat{\mathbf{y}}\}+ (1-p_r){\mathbf{y'}}\{p_r{\sf F}\hat{\mathbf{x}}+(1-p_r){\sf G}\hat{\mathbf{y}}\},\nonumber
\end{eqnarray} 
then one must consider the second order in $\epsilon$ terms to arrive at 
\begin{eqnarray}
	p_r\hat{\mathbf{x}}\{p_r{\sf A}{\mathbf{x'}}+(1-p_r){\sf B}{\mathbf{y'}}\}+ (1-p_r)\hat{\mathbf{y}}\{p_r{\sf F}{\mathbf{x'}}+(1-p_r){\sf G}{\mathbf{y'}}\}
	>
	p_r{\mathbf{x'}}\{p_r{\sf A}{\mathbf{x'}}+(1-p_r){\sf B}{\mathbf{y'}}\}+ (1-p_r){\mathbf{y'}}\{p_r{\sf F}{\mathbf{x'}}+(1-p_r){\sf G}{\mathbf{y'}}\}.\nonumber
\end{eqnarray}
We now relabel $ (\mathbf{x'},\mathbf{y'})$ as $(\mathbf{x}, \mathbf{y})$. Thus, inequality~(\ref{Cond: ESS}) can be equivalently written as a combination of two conditions:  (i) $\forall (\mathbf{x}, \mathbf{y})\neq(\hat{\mathbf{x}},\hat{\mathbf{y}}),$
\begin{eqnarray} 
	p_r\hat{\mathbf{x}}\{p_r{\sf A}\hat{\mathbf{x}}+(1-p_r){\sf B}\hat{\mathbf{y}}\}+ (1-p_r)\hat{\mathbf{y}}\{p_r{\sf F}\hat{\mathbf{x}}+(1-p_r){\sf G}\hat{\mathbf{y}}\}
	\ge
	p_r{\mathbf{x}}\{p_r{\sf A}\hat{\mathbf{x}}+(1-p_r){\sf B}\hat{\mathbf{y}}\}+ (1-p_r){\mathbf{y}}\{p_r{\sf F}\hat{\mathbf{x}}+(1-p_r){\sf G}\hat{\mathbf{y}}\};\nonumber
	\label{Cond: BNE}
\end{eqnarray}
and (ii) $\forall (\mathbf{x}, \mathbf{y})$ for which equality in condition (i) holds, 
\begin{eqnarray}
	p_r\hat{\mathbf{x}}\{p_r{\sf A}{\mathbf{x}}+(1-p_r){\sf B}{\mathbf{y}}\}+ (1-p_r)\hat{\mathbf{y}}\{p_r{\sf F}{\mathbf{x}}+(1-p_r){\sf G}{\mathbf{y}}\}
	>
	p_r{\mathbf{x}}\{p_r{\sf A}{\mathbf{x}}+(1-p_r){\sf B}{\mathbf{y}}\}+ (1-p_r){\mathbf{y}}\{p_r{\sf F}{\mathbf{x}}+(1-p_r){\sf G}{\mathbf{y}}\}.\nonumber
	\label{Cond: stability}
\end{eqnarray}
One notes that condition (i) is nothing but the definition of BNE. Thus,  BESS implies BNE, by construction. 
\end{widetext}

Next, to show BNE---and hence the corresponding BESS---is a fixed point of Eq.~(\ref{eqn: evolution_D_0_matrix_form}), we consider first a BESS that is a weak BNE. A weak BNE, by definition, corresponds to the equality in condition (i). We recast the equality as 
\begin{eqnarray}
	&&\sum_{i}x_ip_r(\mathbf{e}_i-\hat{\mathbf{x}})[p_r{\sf A}\hat{\mathbf{x}}+(1-p_r){\sf B}\hat{\mathbf{y}}]+\nonumber\\&&\phantom{}\sum_{i}y_i(1-p_r)(\mathbf{e}_i-\hat{\mathbf{y}})[p_r{\sf F}\hat{\mathbf{x}}+(1-p_r){\sf G}\hat{\mathbf{y}}]=0\nonumber.
\end{eqnarray}
Since $x_i$ and $y_i$ are the variables that can take arbitrary values and their coefficients do not depend on $x_i$ and $y_i$,  the coefficients must individually vanish. Consequently,  $(\mathbf{e}_i-\mathbf{x})(p_r{\sf A}\mathbf{x}+(1-p_r){\sf B}\mathbf{y})=0$ and $(\mathbf{e}_i-\hat{\mathbf{y}})(p_r{\sf F}\hat{\mathbf{x}}+(1-p_r){\sf G}\hat{\mathbf{y}})=0$, rendering $\dot{x}_i$ and $\dot{y}_i$ in Eq.~(\ref{eqn: evolution_D_0_matrix_form}) to be zero. In other words, the BESS, $(\hat{\mathbf{x}},\hat{\mathbf{y}})$, is a fixed point of Eq.~(\ref{eqn: evolution_D_0_matrix_form}). Finally, if BESS corresponds to a BNE that is strict, then the inequality (without the equal sign) in Eq.~(\ref{Cond: BNE}) should hold by definition and the BESS must be a pure state which must be a vertex of the phase space (a simplotope, in this case). But all vertices are trivially fixed points of Eq.~(\ref{eqn: evolution_D_0_matrix_form}). Therefore, in conclusion, any BESS implies BNE which in turn implies  fixed point.  
\bibliography{patra_etal_reference}

\begin{thebibliography}{67}%
\makeatletter
\providecommand \@ifxundefined [1]{%
 \@ifx{#1\undefined}
}%
\providecommand \@ifnum [1]{%
 \ifnum #1\expandafter \@firstoftwo
 \else \expandafter \@secondoftwo
 \fi
}%
\providecommand \@ifx [1]{%
 \ifx #1\expandafter \@firstoftwo
 \else \expandafter \@secondoftwo
 \fi
}%
\providecommand \natexlab [1]{#1}%
\providecommand \enquote  [1]{``#1''}%
\providecommand \bibnamefont  [1]{#1}%
\providecommand \bibfnamefont [1]{#1}%
\providecommand \citenamefont [1]{#1}%
\providecommand \href@noop [0]{\@secondoftwo}%
\providecommand \href [0]{\begingroup \@sanitize@url \@href}%
\providecommand \@href[1]{\@@startlink{#1}\@@href}%
\providecommand \@@href[1]{\endgroup#1\@@endlink}%
\providecommand \@sanitize@url [0]{\catcode `\\12\catcode `\$12\catcode
  `\&12\catcode `\#12\catcode `\^12\catcode `\_12\catcode `\%12\relax}%
\providecommand \@@startlink[1]{}%
\providecommand \@@endlink[0]{}%
\providecommand \url  [0]{\begingroup\@sanitize@url \@url }%
\providecommand \@url [1]{\endgroup\@href {#1}{\urlprefix }}%
\providecommand \urlprefix  [0]{URL }%
\providecommand \Eprint [0]{\href }%
\providecommand \doibase [0]{https://doi.org/}%
\providecommand \selectlanguage [0]{\@gobble}%
\providecommand \bibinfo  [0]{\@secondoftwo}%
\providecommand \bibfield  [0]{\@secondoftwo}%
\providecommand \translation [1]{[#1]}%
\providecommand \BibitemOpen [0]{}%
\providecommand \bibitemStop [0]{}%
\providecommand \bibitemNoStop [0]{.\EOS\space}%
\providecommand \EOS [0]{\spacefactor3000\relax}%
\providecommand \BibitemShut  [1]{\csname bibitem#1\endcsname}%
\let\auto@bib@innerbib\@empty
\bibitem [{\citenamefont {Weitz}\ \emph {et~al.}(2016)\citenamefont {Weitz},
  \citenamefont {Eksin}, \citenamefont {Paarporn}, \citenamefont {Brown},\ and\
  \citenamefont {Ratcliff}}]{Weitz2016}%
  \BibitemOpen
  \bibfield  {author} {\bibinfo {author} {\bibfnamefont {J.~S.}\ \bibnamefont
  {Weitz}}, \bibinfo {author} {\bibfnamefont {C.}~\bibnamefont {Eksin}},
  \bibinfo {author} {\bibfnamefont {K.}~\bibnamefont {Paarporn}}, \bibinfo
  {author} {\bibfnamefont {S.~P.}\ \bibnamefont {Brown}},\ and\ \bibinfo
  {author} {\bibfnamefont {W.~C.}\ \bibnamefont {Ratcliff}},\ }\bibfield
  {title} {\bibinfo {title} {An oscillating tragedy of the commons in
  replicator dynamics with game-environment feedback},\ }\href
  {http://dx.doi.org/10.1073/pnas.1604096113} {\bibfield  {journal} {\bibinfo
  {journal} {Proc. Natl. Acad. Sci. U.S.A}\ }\textbf {\bibinfo {volume}
  {113}},\ \bibinfo {pages} {E7518–E7525} (\bibinfo {year}
  {2016})}\BibitemShut {NoStop}%
\bibitem [{\citenamefont {Tilman}\ \emph {et~al.}(2020)\citenamefont {Tilman},
  \citenamefont {Plotkin},\ and\ \citenamefont {Ak\c{c}ay}}]{Tilman2020}%
  \BibitemOpen
  \bibfield  {author} {\bibinfo {author} {\bibfnamefont {A.~R.}\ \bibnamefont
  {Tilman}}, \bibinfo {author} {\bibfnamefont {J.~B.}\ \bibnamefont
  {Plotkin}},\ and\ \bibinfo {author} {\bibfnamefont {E.}~\bibnamefont
  {Ak\c{c}ay}},\ }\bibfield  {title} {\bibinfo {title} {Evolutionary games with
  environmental feedbacks},\ }\href
  {http://dx.doi.org/10.1038/s41467-020-14531-6} {\bibfield  {journal}
  {\bibinfo  {journal} {Nat. Commun.}\ }\textbf {\bibinfo {volume} {11}},\
  \bibinfo {pages} {915} (\bibinfo {year} {2020})}\BibitemShut {NoStop}%
\bibitem [{\citenamefont {Lin}\ and\ \citenamefont {Weitz}(2019)}]{Lin2019}%
  \BibitemOpen
  \bibfield  {author} {\bibinfo {author} {\bibfnamefont {Y.-H.}\ \bibnamefont
  {Lin}}\ and\ \bibinfo {author} {\bibfnamefont {J.~S.}\ \bibnamefont
  {Weitz}},\ }\bibfield  {title} {\bibinfo {title} {Spatial interactions and
  oscillatory tragedies of the commons},\ }\href
  {http://dx.doi.org/10.1103/PhysRevLett.122.148102} {\bibfield  {journal}
  {\bibinfo  {journal} {Phys. Rev. Lett.}\ }\textbf {\bibinfo {volume} {122}},\
  \bibinfo {pages} {148102} (\bibinfo {year} {2019})}\BibitemShut {NoStop}%
\bibitem [{\citenamefont {Harsanyi}(1982)}]{Harsanyi1982}%
  \BibitemOpen
  \bibfield  {author} {\bibinfo {author} {\bibfnamefont {J.~C.}\ \bibnamefont
  {Harsanyi}},\ }\href {https://doi.org/10.1007/978-94-017-2527-9} {\emph
  {\bibinfo {title} {Papers in Game Theory}}}\ (\bibinfo  {publisher} {Springer
  Netherlands},\ \bibinfo {year} {1982})\BibitemShut {NoStop}%
\bibitem [{\citenamefont {Bonanno}(2018)}]{Bonanno2018}%
  \BibitemOpen
  \bibfield  {author} {\bibinfo {author} {\bibfnamefont {G.}~\bibnamefont
  {Bonanno}},\ }\href@noop {} {\emph {\bibinfo {title} {Game theory}}}\
  (\bibinfo  {publisher} {CreateSpace Independent Publishing Platform},\
  \bibinfo {year} {2018})\BibitemShut {NoStop}%
\bibitem [{\citenamefont {Li}(1985)}]{Li1985}%
  \BibitemOpen
  \bibfield  {author} {\bibinfo {author} {\bibfnamefont {L.}~\bibnamefont
  {Li}},\ }\bibfield  {title} {\bibinfo {title} {Cournot oligopoly with
  information sharing},\ }\href {https://www.jstor.org/stable/2555510}
  {\bibfield  {journal} {\bibinfo  {journal} {RAND J. Econ.}\ }\textbf
  {\bibinfo {volume} {16}},\ \bibinfo {pages} {521} (\bibinfo {year}
  {1985})}\BibitemShut {NoStop}%
\bibitem [{\citenamefont {Jin}\ \emph {et~al.}(2013)\citenamefont {Jin},
  \citenamefont {Pissinou}, \citenamefont {Pumpichet}, \citenamefont
  {Kamhoua},\ and\ \citenamefont {Kwiat}}]{XinyuJin2013}%
  \BibitemOpen
  \bibfield  {author} {\bibinfo {author} {\bibfnamefont {X.}~\bibnamefont
  {Jin}}, \bibinfo {author} {\bibfnamefont {N.}~\bibnamefont {Pissinou}},
  \bibinfo {author} {\bibfnamefont {S.}~\bibnamefont {Pumpichet}}, \bibinfo
  {author} {\bibfnamefont {C.~A.}\ \bibnamefont {Kamhoua}},\ and\ \bibinfo
  {author} {\bibfnamefont {K.}~\bibnamefont {Kwiat}},\ }\bibfield  {title}
  {\bibinfo {title} {Modeling cooperative, selfish and malicious behaviors for
  trajectory privacy preservation using bayesian game theory},\ }\href
  {http://dx.doi.org/10.1109/LCN.2013.6761339} {\bibfield  {journal} {\bibinfo
  {journal} {38th Annual IEEE Conference on Local Computer Networks}\ }\textbf
  {\bibinfo {volume} {2}},\ \bibinfo {pages} {835} (\bibinfo {year}
  {2013})}\BibitemShut {NoStop}%
\bibitem [{\citenamefont {Mohi}\ \emph {et~al.}(2009)\citenamefont {Mohi},
  \citenamefont {Movaghar},\ and\ \citenamefont {Zadeh}}]{Mohi2009}%
  \BibitemOpen
  \bibfield  {author} {\bibinfo {author} {\bibfnamefont {M.}~\bibnamefont
  {Mohi}}, \bibinfo {author} {\bibfnamefont {A.}~\bibnamefont {Movaghar}},\
  and\ \bibinfo {author} {\bibfnamefont {P.~M.}\ \bibnamefont {Zadeh}},\
  }\bibfield  {title} {\bibinfo {title} {A bayesian game approach for
  preventing dos attacks in wireless sensor networks},\ }\href
  {https://doi.org/10.1109/CMC.2009.325} {\bibfield  {journal} {\bibinfo
  {journal} {2009 WRI International Conference on Communications and Mobile
  Computing}\ }\textbf {\bibinfo {volume} {3}},\ \bibinfo {pages} {507}
  (\bibinfo {year} {2009})}\BibitemShut {NoStop}%
\bibitem [{\citenamefont {Smith}\ and\ \citenamefont
  {Price}(1973)}]{Smith1973}%
  \BibitemOpen
  \bibfield  {author} {\bibinfo {author} {\bibfnamefont {J.~M.}\ \bibnamefont
  {Smith}}\ and\ \bibinfo {author} {\bibfnamefont {G.~R.}\ \bibnamefont
  {Price}},\ }\bibfield  {title} {\bibinfo {title} {The logic of animal
  conflict},\ }\href {https://doi.org/10.1038/246015a0} {\bibfield  {journal}
  {\bibinfo  {journal} {Nature}\ }\textbf {\bibinfo {volume} {246}},\ \bibinfo
  {pages} {15–18} (\bibinfo {year} {1973})}\BibitemShut {NoStop}%
\bibitem [{\citenamefont {Maynard~Smith}(1974)}]{MaynardSmith1974}%
  \BibitemOpen
  \bibfield  {author} {\bibinfo {author} {\bibfnamefont {J.}~\bibnamefont
  {Maynard~Smith}},\ }\bibfield  {title} {\bibinfo {title} {The theory of games
  and the evolution of animal conflicts},\ }\href
  {https://doi.org/10.1016/0022-5193(74)90110-6} {\bibfield  {journal}
  {\bibinfo  {journal} {J. Theor. Biol.}\ }\textbf {\bibinfo {volume} {47}},\
  \bibinfo {pages} {209–221} (\bibinfo {year} {1974})}\BibitemShut {NoStop}%
\bibitem [{\citenamefont {Smith}(1988)}]{Smith1988}%
  \BibitemOpen
  \bibfield  {author} {\bibinfo {author} {\bibfnamefont {J.~M.}\ \bibnamefont
  {Smith}},\ }\href {https://doi.org/10.1007/978-1-4684-7862-4_22} {\emph
  {\bibinfo {title} {Did Darwin Get It Right?}}}\ (\bibinfo  {publisher}
  {Springer US},\ \bibinfo {year} {1988})\ p.\ \bibinfo {pages}
  {202–215}\BibitemShut {NoStop}%
\bibitem [{\citenamefont {Hardin}(1968)}]{Hardin1968}%
  \BibitemOpen
  \bibfield  {author} {\bibinfo {author} {\bibfnamefont {G.}~\bibnamefont
  {Hardin}},\ }\bibfield  {title} {\bibinfo {title} {The tragedy of the
  commons: The population problem has no technical solution; it requires a
  fundamental extension in morality.},\ }\href
  {https://doi.org/10.1126/science.162.3859.1243} {\bibfield  {journal}
  {\bibinfo  {journal} {Science}\ }\textbf {\bibinfo {volume} {162}},\ \bibinfo
  {pages} {1243–1248} (\bibinfo {year} {1968})}\BibitemShut {NoStop}%
\bibitem [{\citenamefont {Ostrom}(1999)}]{Ostrom1999}%
  \BibitemOpen
  \bibfield  {author} {\bibinfo {author} {\bibfnamefont {E.}~\bibnamefont
  {Ostrom}},\ }\bibfield  {title} {\bibinfo {title} {Coping with tragedies of
  the commons},\ }\href {https://doi.org/10.1146/annurev.polisci.2.1.493}
  {\bibfield  {journal} {\bibinfo  {journal} {Annu. Rev. Political Sci.}\
  }\textbf {\bibinfo {volume} {2}},\ \bibinfo {pages} {493–535} (\bibinfo
  {year} {1999})}\BibitemShut {NoStop}%
\bibitem [{\citenamefont {Mondal}\ \emph {et~al.}(2022)\citenamefont {Mondal},
  \citenamefont {Pathak},\ and\ \citenamefont {Chakraborty}}]{Mondal2022}%
  \BibitemOpen
  \bibfield  {author} {\bibinfo {author} {\bibfnamefont {S.~S.}\ \bibnamefont
  {Mondal}}, \bibinfo {author} {\bibfnamefont {M.}~\bibnamefont {Pathak}},\
  and\ \bibinfo {author} {\bibfnamefont {S.}~\bibnamefont {Chakraborty}},\
  }\bibfield  {title} {\bibinfo {title} {Reward versus punishment: averting the
  tragedy of the commons in eco-evolutionary dynamics},\ }\href
  {https://doi.org/10.1088/2632-072x/ac6c6e} {\bibfield  {journal} {\bibinfo
  {journal} {J. Phys.: Complex.}\ }\textbf {\bibinfo {volume} {3}},\ \bibinfo
  {pages} {025005} (\bibinfo {year} {2022})}\BibitemShut {NoStop}%
\bibitem [{\citenamefont {Das~Bairagya}\ \emph {et~al.}(2021)\citenamefont
  {Das~Bairagya}, \citenamefont {Mondal}, \citenamefont {Chowdhury},\ and\
  \citenamefont {Chakraborty}}]{DasBairagya2021}%
  \BibitemOpen
  \bibfield  {author} {\bibinfo {author} {\bibfnamefont {J.}~\bibnamefont
  {Das~Bairagya}}, \bibinfo {author} {\bibfnamefont {S.~S.}\ \bibnamefont
  {Mondal}}, \bibinfo {author} {\bibfnamefont {D.}~\bibnamefont {Chowdhury}},\
  and\ \bibinfo {author} {\bibfnamefont {S.}~\bibnamefont {Chakraborty}},\
  }\bibfield  {title} {\bibinfo {title} {Game-environment feedback dynamics in
  growing population: Effect of finite carrying capacity},\ }\href
  {http://dx.doi.org/10.1103/PhysRevE.104.044407} {\bibfield  {journal}
  {\bibinfo  {journal} {Phys. Rev. E}\ }\textbf {\bibinfo {volume} {104}},\
  \bibinfo {pages} {044407} (\bibinfo {year} {2021})}\BibitemShut {NoStop}%
\bibitem [{\citenamefont {Sohel~Mondal}\ \emph {et~al.}(2024)\citenamefont
  {Sohel~Mondal}, \citenamefont {Ray},\ and\ \citenamefont
  {Chakraborty}}]{SohelMondal2024}%
  \BibitemOpen
  \bibfield  {author} {\bibinfo {author} {\bibfnamefont {S.}~\bibnamefont
  {Sohel~Mondal}}, \bibinfo {author} {\bibfnamefont {A.}~\bibnamefont {Ray}},\
  and\ \bibinfo {author} {\bibfnamefont {S.}~\bibnamefont {Chakraborty}},\
  }\bibfield  {title} {\bibinfo {title} {Hypochaos prevents tragedy of the
  commons in discrete-time eco-evolutionary game dynamics},\ }\href
  {http://dx.doi.org/10.1063/5.0190800} {\bibfield  {journal} {\bibinfo
  {journal} {Chaos}\ }\textbf {\bibinfo {volume} {34}},\ \bibinfo {pages}
  {023122} (\bibinfo {year} {2024})}\BibitemShut {NoStop}%
\bibitem [{\citenamefont {Bairagya}\ \emph {et~al.}(2023)\citenamefont
  {Bairagya}, \citenamefont {Mondal}, \citenamefont {Chowdhury},\ and\
  \citenamefont {Chakraborty}}]{Bairagya2023}%
  \BibitemOpen
  \bibfield  {author} {\bibinfo {author} {\bibfnamefont {J.~D.}\ \bibnamefont
  {Bairagya}}, \bibinfo {author} {\bibfnamefont {S.~S.}\ \bibnamefont
  {Mondal}}, \bibinfo {author} {\bibfnamefont {D.}~\bibnamefont {Chowdhury}},\
  and\ \bibinfo {author} {\bibfnamefont {S.}~\bibnamefont {Chakraborty}},\
  }\bibfield  {title} {\bibinfo {title} {Eco-evolutionary games for harvesting
  self-renewing common resource: effect of growing harvester population},\
  }\href {https://doi.org/10.1088/2632-072x/acc5cb} {\bibfield  {journal}
  {\bibinfo  {journal} {J. Phys.: Complex.}\ }\textbf {\bibinfo {volume} {4}},\
  \bibinfo {pages} {025002} (\bibinfo {year} {2023})}\BibitemShut {NoStop}%
\bibitem [{\citenamefont {Roy}\ \emph {et~al.}(2023)\citenamefont {Roy},
  \citenamefont {Nag~Chowdhury}, \citenamefont {Kundu}, \citenamefont {Sar},
  \citenamefont {Banerjee}, \citenamefont {Rakshit}, \citenamefont {Mali},
  \citenamefont {Perc},\ and\ \citenamefont {Ghosh}}]{Roy2023}%
  \BibitemOpen
  \bibfield  {author} {\bibinfo {author} {\bibfnamefont {S.}~\bibnamefont
  {Roy}}, \bibinfo {author} {\bibfnamefont {S.}~\bibnamefont {Nag~Chowdhury}},
  \bibinfo {author} {\bibfnamefont {S.}~\bibnamefont {Kundu}}, \bibinfo
  {author} {\bibfnamefont {G.~K.}\ \bibnamefont {Sar}}, \bibinfo {author}
  {\bibfnamefont {J.}~\bibnamefont {Banerjee}}, \bibinfo {author}
  {\bibfnamefont {B.}~\bibnamefont {Rakshit}}, \bibinfo {author} {\bibfnamefont
  {P.~C.}\ \bibnamefont {Mali}}, \bibinfo {author} {\bibfnamefont
  {M.}~\bibnamefont {Perc}},\ and\ \bibinfo {author} {\bibfnamefont
  {D.}~\bibnamefont {Ghosh}},\ }\bibfield  {title} {\bibinfo {title} {Time
  delays shape the eco-evolutionary dynamics of cooperation},\ }\href
  {http://dx.doi.org/10.1038/s41598-023-41519-1} {\bibfield  {journal}
  {\bibinfo  {journal} {Sci. Rep.}\ }\textbf {\bibinfo {volume} {13}},\
  \bibinfo {pages} {14331} (\bibinfo {year} {2023})}\BibitemShut {NoStop}%
\bibitem [{\citenamefont {Roy}\ \emph {et~al.}(2024)\citenamefont {Roy},
  \citenamefont {Ghosh}, \citenamefont {Saha}, \citenamefont {Chandra~Mali},
  \citenamefont {Perc},\ and\ \citenamefont {Ghosh}}]{Roy2024}%
  \BibitemOpen
  \bibfield  {author} {\bibinfo {author} {\bibfnamefont {S.}~\bibnamefont
  {Roy}}, \bibinfo {author} {\bibfnamefont {S.}~\bibnamefont {Ghosh}}, \bibinfo
  {author} {\bibfnamefont {A.}~\bibnamefont {Saha}}, \bibinfo {author}
  {\bibfnamefont {P.}~\bibnamefont {Chandra~Mali}}, \bibinfo {author}
  {\bibfnamefont {M.}~\bibnamefont {Perc}},\ and\ \bibinfo {author}
  {\bibfnamefont {D.}~\bibnamefont {Ghosh}},\ }\bibfield  {title} {\bibinfo
  {title} {The eco-evolutionary dynamics of strategic species},\ }\href
  {http://dx.doi.org/10.1098/rspa.2024.0127} {\bibfield  {journal} {\bibinfo
  {journal} {Proc. R. Soc. A: Math. Phys. Eng. Sci.}\ }\textbf {\bibinfo
  {volume} {480}},\ \bibinfo {pages} {20240127} (\bibinfo {year}
  {2024})}\BibitemShut {NoStop}%
\bibitem [{\citenamefont {Ghosh}\ \emph {et~al.}(2024)\citenamefont {Ghosh},
  \citenamefont {Roy}, \citenamefont {Perc},\ and\ \citenamefont
  {Ghosh}}]{Ghosh2024}%
  \BibitemOpen
  \bibfield  {author} {\bibinfo {author} {\bibfnamefont {S.}~\bibnamefont
  {Ghosh}}, \bibinfo {author} {\bibfnamefont {S.}~\bibnamefont {Roy}}, \bibinfo
  {author} {\bibfnamefont {M.}~\bibnamefont {Perc}},\ and\ \bibinfo {author}
  {\bibfnamefont {D.}~\bibnamefont {Ghosh}},\ }\bibfield  {title} {\bibinfo
  {title} {The eco-evolutionary dynamics of two strategic species: From the
  predator-prey to the innocent-spreader rumor model},\ }\href
  {https://doi.org/10.1016/j.jtbi.2024.111955} {\bibfield  {journal} {\bibinfo
  {journal} {J. Theor. Biol.}\ }\textbf {\bibinfo {volume} {595}},\ \bibinfo
  {pages} {111955} (\bibinfo {year} {2024})}\BibitemShut {NoStop}%
\bibitem [{\citenamefont {Szabó}\ and\ \citenamefont
  {Fáth}(2007)}]{Szab2007}%
  \BibitemOpen
  \bibfield  {author} {\bibinfo {author} {\bibfnamefont {G.}~\bibnamefont
  {Szabó}}\ and\ \bibinfo {author} {\bibfnamefont {G.}~\bibnamefont {Fáth}},\
  }\bibfield  {title} {\bibinfo {title} {Evolutionary games on graphs},\ }\href
  {https://doi.org/10.1016/j.physrep.2007.04.004} {\bibfield  {journal}
  {\bibinfo  {journal} {Phys. Rep.}\ }\textbf {\bibinfo {volume} {446}},\
  \bibinfo {pages} {97–216} (\bibinfo {year} {2007})}\BibitemShut {NoStop}%
\bibitem [{\citenamefont {Perc}\ and\ \citenamefont
  {Szolnoki}(2010)}]{Perc2010}%
  \BibitemOpen
  \bibfield  {author} {\bibinfo {author} {\bibfnamefont {M.}~\bibnamefont
  {Perc}}\ and\ \bibinfo {author} {\bibfnamefont {A.}~\bibnamefont
  {Szolnoki}},\ }\bibfield  {title} {\bibinfo {title} {Coevolutionary games—a
  mini review},\ }\href {https://doi.org/10.1016/j.biosystems.2009.10.003}
  {\bibfield  {journal} {\bibinfo  {journal} {Biosystems}\ }\textbf {\bibinfo
  {volume} {99}},\ \bibinfo {pages} {109–125} (\bibinfo {year}
  {2010})}\BibitemShut {NoStop}%
\bibitem [{\citenamefont {Perc}(2006{\natexlab{a}})}]{Perc2006a}%
  \BibitemOpen
  \bibfield  {author} {\bibinfo {author} {\bibfnamefont {M.}~\bibnamefont
  {Perc}},\ }\bibfield  {title} {\bibinfo {title} {Chaos promotes cooperation
  in the spatial prisoner’s dilemma game},\ }\href
  {https://doi.org/10.1209/epl/i2006-10217-3} {\bibfield  {journal} {\bibinfo
  {journal} {Europhys. Lett.}\ }\textbf {\bibinfo {volume} {75}},\ \bibinfo
  {pages} {841–846} (\bibinfo {year} {2006}{\natexlab{a}})}\BibitemShut
  {NoStop}%
\bibitem [{\citenamefont {Perc}(2006{\natexlab{b}})}]{Perc2006b}%
  \BibitemOpen
  \bibfield  {author} {\bibinfo {author} {\bibfnamefont {M.}~\bibnamefont
  {Perc}},\ }\bibfield  {title} {\bibinfo {title} {Coherence resonance in a
  spatial prisoner’s dilemma game},\ }\href
  {https://doi.org/10.1088/1367-2630/8/2/022} {\bibfield  {journal} {\bibinfo
  {journal} {New J. Phys.}\ }\textbf {\bibinfo {volume} {8}},\ \bibinfo {pages}
  {22–22} (\bibinfo {year} {2006}{\natexlab{b}})}\BibitemShut {NoStop}%
\bibitem [{\citenamefont {Taylor}\ and\ \citenamefont
  {Jonker}(1978)}]{Taylor1978}%
  \BibitemOpen
  \bibfield  {author} {\bibinfo {author} {\bibfnamefont {P.~D.}\ \bibnamefont
  {Taylor}}\ and\ \bibinfo {author} {\bibfnamefont {L.~B.}\ \bibnamefont
  {Jonker}},\ }\bibfield  {title} {\bibinfo {title} {Evolutionary stable
  strategies and game dynamics},\ }\href
  {https://doi.org/10.1016/0025-5564(78)90077-9} {\bibfield  {journal}
  {\bibinfo  {journal} {Math. Biosci.}\ }\textbf {\bibinfo {volume} {40}},\
  \bibinfo {pages} {145–156} (\bibinfo {year} {1978})}\BibitemShut {NoStop}%
\bibitem [{\citenamefont {Schuster}\ and\ \citenamefont
  {Sigmund}(1983)}]{Schuster1983}%
  \BibitemOpen
  \bibfield  {author} {\bibinfo {author} {\bibfnamefont {P.}~\bibnamefont
  {Schuster}}\ and\ \bibinfo {author} {\bibfnamefont {K.}~\bibnamefont
  {Sigmund}},\ }\bibfield  {title} {\bibinfo {title} {Replicator dynamics},\
  }\href {https://doi.org/10.1016/0022-5193(83)90445-9} {\bibfield  {journal}
  {\bibinfo  {journal} {J. Theor. Biol.}\ }\textbf {\bibinfo {volume} {100}},\
  \bibinfo {pages} {533–538} (\bibinfo {year} {1983})}\BibitemShut {NoStop}%
\bibitem [{\citenamefont {Cressman}\ and\ \citenamefont
  {Tao}(2014)}]{Cressman2014}%
  \BibitemOpen
  \bibfield  {author} {\bibinfo {author} {\bibfnamefont {R.}~\bibnamefont
  {Cressman}}\ and\ \bibinfo {author} {\bibfnamefont {Y.}~\bibnamefont {Tao}},\
  }\bibfield  {title} {\bibinfo {title} {The replicator equation and other game
  dynamics},\ }\href {https://doi.org/10.1073/pnas.1400823111} {\bibfield
  {journal} {\bibinfo  {journal} {Proc. Natl. Acad. Sci. U.S.A}\ }\textbf
  {\bibinfo {volume} {111}},\ \bibinfo {pages} {10810–10817} (\bibinfo {year}
  {2014})}\BibitemShut {NoStop}%
\bibitem [{\citenamefont {May}(1975)}]{May1975}%
  \BibitemOpen
  \bibfield  {author} {\bibinfo {author} {\bibfnamefont {R.~M.}\ \bibnamefont
  {May}},\ }\bibfield  {title} {\bibinfo {title} {Biological populations
  obeying difference equations: Stable points, stable cycles, and chaos},\
  }\href {https://doi.org/10.1016/0022-5193(75)90078-8} {\bibfield  {journal}
  {\bibinfo  {journal} {J. Theor. Biol.}\ }\textbf {\bibinfo {volume} {51}},\
  \bibinfo {pages} {511–524} (\bibinfo {year} {1975})}\BibitemShut {NoStop}%
\bibitem [{\citenamefont {Gong}\ \emph {et~al.}(2022)\citenamefont {Gong},
  \citenamefont {Yao}, \citenamefont {Gao},\ and\ \citenamefont
  {Cao}}]{Gong2022}%
  \BibitemOpen
  \bibfield  {author} {\bibinfo {author} {\bibfnamefont {L.}~\bibnamefont
  {Gong}}, \bibinfo {author} {\bibfnamefont {W.}~\bibnamefont {Yao}}, \bibinfo
  {author} {\bibfnamefont {J.}~\bibnamefont {Gao}},\ and\ \bibinfo {author}
  {\bibfnamefont {M.}~\bibnamefont {Cao}},\ }\bibfield  {title} {\bibinfo
  {title} {Limit cycles analysis and control of evolutionary game dynamics with
  environmental feedback},\ }\href
  {https://doi.org/10.1016/j.automatica.2022.110536} {\bibfield  {journal}
  {\bibinfo  {journal} {Automatica}\ }\textbf {\bibinfo {volume} {145}},\
  \bibinfo {pages} {110536} (\bibinfo {year} {2022})}\BibitemShut {NoStop}%
\bibitem [{\citenamefont {Rivoire}\ and\ \citenamefont
  {Leibler}(2011)}]{Rivoire2011}%
  \BibitemOpen
  \bibfield  {author} {\bibinfo {author} {\bibfnamefont {O.}~\bibnamefont
  {Rivoire}}\ and\ \bibinfo {author} {\bibfnamefont {S.}~\bibnamefont
  {Leibler}},\ }\bibfield  {title} {\bibinfo {title} {The value of information
  for populations in varying environments},\ }\href
  {https://doi.org/10.1007/s10955-011-0166-2} {\bibfield  {journal} {\bibinfo
  {journal} {J. Stat. Phys.}\ }\textbf {\bibinfo {volume} {142}},\ \bibinfo
  {pages} {1124–1166} (\bibinfo {year} {2011})}\BibitemShut {NoStop}%
\bibitem [{\citenamefont {Barfuss}\ and\ \citenamefont
  {Mann}(2022)}]{Barfuss2022}%
  \BibitemOpen
  \bibfield  {author} {\bibinfo {author} {\bibfnamefont {W.}~\bibnamefont
  {Barfuss}}\ and\ \bibinfo {author} {\bibfnamefont {R.~P.}\ \bibnamefont
  {Mann}},\ }\bibfield  {title} {\bibinfo {title} {Modeling the effects of
  environmental and perceptual uncertainty using deterministic reinforcement
  learning dynamics with partial observability},\ }\href
  {http://dx.doi.org/10.1103/PhysRevE.105.034409} {\bibfield  {journal}
  {\bibinfo  {journal} {Phys. Rev. E}\ }\textbf {\bibinfo {volume} {105}},\
  \bibinfo {pages} {034409} (\bibinfo {year} {2022})}\BibitemShut {NoStop}%
\bibitem [{\citenamefont {Kleshnina}\ \emph {et~al.}(2023)\citenamefont
  {Kleshnina}, \citenamefont {Hilbe}, \citenamefont {Simsa}, \citenamefont
  {Chatterjee},\ and\ \citenamefont {Nowak}}]{Kleshnina2023}%
  \BibitemOpen
  \bibfield  {author} {\bibinfo {author} {\bibfnamefont {M.}~\bibnamefont
  {Kleshnina}}, \bibinfo {author} {\bibfnamefont {C.}~\bibnamefont {Hilbe}},
  \bibinfo {author} {\bibfnamefont {S.}~\bibnamefont {Simsa}}, \bibinfo
  {author} {\bibfnamefont {K.}~\bibnamefont {Chatterjee}},\ and\ \bibinfo
  {author} {\bibfnamefont {M.~A.}\ \bibnamefont {Nowak}},\ }\bibfield  {title}
  {\bibinfo {title} {The effect of environmental information on evolution of
  cooperation in stochastic games},\ }\href
  {http://dx.doi.org/10.1038/s41467-023-39625-9} {\bibfield  {journal}
  {\bibinfo  {journal} {Nat. Commun.}\ }\textbf {\bibinfo {volume} {14}},\
  \bibinfo {pages} {4153} (\bibinfo {year} {2023})}\BibitemShut {NoStop}%
\bibitem [{\citenamefont {Slatkin}(1974)}]{Slatkin1974}%
  \BibitemOpen
  \bibfield  {author} {\bibinfo {author} {\bibfnamefont {M.}~\bibnamefont
  {Slatkin}},\ }\bibfield  {title} {\bibinfo {title} {Hedging one’s
  evolutionary bets},\ }\href {https://doi.org/10.1038/250704b0} {\bibfield
  {journal} {\bibinfo  {journal} {Nature}\ }\textbf {\bibinfo {volume} {250}},\
  \bibinfo {pages} {704–705} (\bibinfo {year} {1974})}\BibitemShut {NoStop}%
\bibitem [{\citenamefont {Harvey}\ and\ \citenamefont
  {Partridge}(1987)}]{Harvey1987}%
  \BibitemOpen
  \bibfield  {author} {\bibinfo {author} {\bibfnamefont {P.~H.}\ \bibnamefont
  {Harvey}}\ and\ \bibinfo {author} {\bibfnamefont {L.}~\bibnamefont
  {Partridge}},\ }\href@noop {} {\emph {\bibinfo {title} {Oxford Surveys in
  Evolutionary Biology}}},\ Vol.~\bibinfo {volume} {4}\ (\bibinfo  {publisher}
  {Oxford University Press},\ \bibinfo {year} {1987})\BibitemShut {NoStop}%
\bibitem [{\citenamefont {Gillespie}(1974)}]{Gillespie1974}%
  \BibitemOpen
  \bibfield  {author} {\bibinfo {author} {\bibfnamefont {J.~H.}\ \bibnamefont
  {Gillespie}},\ }\bibfield  {title} {\bibinfo {title} {Natural selection for
  within-generation variance in offspring number},\ }\href
  {https://doi.org/10.1093/genetics/76.3.601} {\bibfield  {journal} {\bibinfo
  {journal} {Genetics}\ }\textbf {\bibinfo {volume} {76}},\ \bibinfo {pages}
  {601–606} (\bibinfo {year} {1974})}\BibitemShut {NoStop}%
\bibitem [{\citenamefont {Einum}\ and\ \citenamefont
  {Fleming}(1999)}]{Einum1999}%
  \BibitemOpen
  \bibfield  {author} {\bibinfo {author} {\bibfnamefont {S.}~\bibnamefont
  {Einum}}\ and\ \bibinfo {author} {\bibfnamefont {I.}~\bibnamefont
  {Fleming}},\ }\bibfield  {title} {\bibinfo {title} {Maternal effects of egg
  size in brown trout (salmo trutta): norms of reaction to environmental
  quality},\ }\href {https://doi.org/10.1098/rspb.1999.0893} {\bibfield
  {journal} {\bibinfo  {journal} {Proc. R. Soc. B: Biol. Sci.}\ }\textbf
  {\bibinfo {volume} {266}},\ \bibinfo {pages} {2095–2100} (\bibinfo {year}
  {1999})}\BibitemShut {NoStop}%
\bibitem [{\citenamefont {Cohen}(1966)}]{Cohen1966}%
  \BibitemOpen
  \bibfield  {author} {\bibinfo {author} {\bibfnamefont {D.}~\bibnamefont
  {Cohen}},\ }\bibfield  {title} {\bibinfo {title} {Optimizing reproduction in
  a randomly varying environment},\ }\href
  {https://doi.org/10.1016/0022-5193(66)90188-3} {\bibfield  {journal}
  {\bibinfo  {journal} {J. Theor. Biol.}\ }\textbf {\bibinfo {volume} {12}},\
  \bibinfo {pages} {119–129} (\bibinfo {year} {1966})}\BibitemShut {NoStop}%
\bibitem [{\citenamefont {Cooper}\ and\ \citenamefont
  {Kaplan}(1982)}]{Cooper1982}%
  \BibitemOpen
  \bibfield  {author} {\bibinfo {author} {\bibfnamefont {W.~S.}\ \bibnamefont
  {Cooper}}\ and\ \bibinfo {author} {\bibfnamefont {R.~H.}\ \bibnamefont
  {Kaplan}},\ }\bibfield  {title} {\bibinfo {title} {Adaptive
  “coin-flipping”: a decision-theoretic examination of natural selection
  for random individual variation},\ }\href
  {https://doi.org/10.1016/0022-5193(82)90336-8} {\bibfield  {journal}
  {\bibinfo  {journal} {J. Theor. Biol.}\ }\textbf {\bibinfo {volume} {94}},\
  \bibinfo {pages} {135–151} (\bibinfo {year} {1982})}\BibitemShut {NoStop}%
\bibitem [{\citenamefont {Shannon}(1948)}]{Shannon1948}%
  \BibitemOpen
  \bibfield  {author} {\bibinfo {author} {\bibfnamefont {C.~E.}\ \bibnamefont
  {Shannon}},\ }\bibfield  {title} {\bibinfo {title} {A mathematical theory of
  communication},\ }\href {https://doi.org/10.1002/j.1538-7305.1948.tb01338.x}
  {\bibfield  {journal} {\bibinfo  {journal} {Bell Syst. Tech. J.}\ }\textbf
  {\bibinfo {volume} {27}},\ \bibinfo {pages} {379–423} (\bibinfo {year}
  {1948})}\BibitemShut {NoStop}%
\bibitem [{\citenamefont {Rankin}\ \emph {et~al.}(2007)\citenamefont {Rankin},
  \citenamefont {Bargum},\ and\ \citenamefont {Kokko}}]{Rankin2007}%
  \BibitemOpen
  \bibfield  {author} {\bibinfo {author} {\bibfnamefont {D.~J.}\ \bibnamefont
  {Rankin}}, \bibinfo {author} {\bibfnamefont {K.}~\bibnamefont {Bargum}},\
  and\ \bibinfo {author} {\bibfnamefont {H.}~\bibnamefont {Kokko}},\ }\bibfield
   {title} {\bibinfo {title} {The tragedy of the commons in evolutionary
  biology},\ }\href {https://doi.org/10.1016/j.tree.2007.07.009} {\bibfield
  {journal} {\bibinfo  {journal} {Trends Ecol. Evol.}\ }\textbf {\bibinfo
  {volume} {22}},\ \bibinfo {pages} {643–651} (\bibinfo {year}
  {2007})}\BibitemShut {NoStop}%
\bibitem [{\citenamefont {Keynes}(2013)}]{keynes2013treatise}%
  \BibitemOpen
  \bibfield  {author} {\bibinfo {author} {\bibfnamefont {J.~M.}\ \bibnamefont
  {Keynes}},\ }\href@noop {} {\emph {\bibinfo {title} {A treatise on
  probability}}}\ (\bibinfo  {publisher} {Courier Corporation},\ \bibinfo
  {year} {2013})\BibitemShut {NoStop}%
\bibitem [{\citenamefont {Bodmer}\ and\ \citenamefont
  {Felsenstein}(1967)}]{Bodmer1967}%
  \BibitemOpen
  \bibfield  {author} {\bibinfo {author} {\bibfnamefont {W.~F.}\ \bibnamefont
  {Bodmer}}\ and\ \bibinfo {author} {\bibfnamefont {J.}~\bibnamefont
  {Felsenstein}},\ }\bibfield  {title} {\bibinfo {title} {Linkage and
  selection: Theoretical analysis of the deterministic two locus random mating
  model},\ }\href {https://doi.org/10.1093/genetics/57.2.237} {\bibfield
  {journal} {\bibinfo  {journal} {Genetics}\ }\textbf {\bibinfo {volume}
  {57}},\ \bibinfo {pages} {237–265} (\bibinfo {year} {1967})}\BibitemShut
  {NoStop}%
\bibitem [{\citenamefont {Ewens}(1968)}]{Ewens1968}%
  \BibitemOpen
  \bibfield  {author} {\bibinfo {author} {\bibfnamefont {W.~J.}\ \bibnamefont
  {Ewens}},\ }\bibfield  {title} {\bibinfo {title} {A genetic model having
  complex linkage behaviour},\ }\href {https://doi.org/10.1007/bf00933808}
  {\bibfield  {journal} {\bibinfo  {journal} {Theor. Appl. Genet.}\ }\textbf
  {\bibinfo {volume} {38}},\ \bibinfo {pages} {140–143} (\bibinfo {year}
  {1968})}\BibitemShut {NoStop}%
\bibitem [{\citenamefont {Karlin}\ and\ \citenamefont
  {Feldman}(1970)}]{Karlin1970}%
  \BibitemOpen
  \bibfield  {author} {\bibinfo {author} {\bibfnamefont {S.}~\bibnamefont
  {Karlin}}\ and\ \bibinfo {author} {\bibfnamefont {M.~W.}\ \bibnamefont
  {Feldman}},\ }\bibfield  {title} {\bibinfo {title} {Linkage and selection:
  Two locus symmetric viability model},\ }\href
  {https://doi.org/10.1016/0040-5809(70)90041-9} {\bibfield  {journal}
  {\bibinfo  {journal} {Theor. Popul. Biol.}\ }\textbf {\bibinfo {volume}
  {1}},\ \bibinfo {pages} {39–71} (\bibinfo {year} {1970})}\BibitemShut
  {NoStop}%
\bibitem [{\citenamefont {Feldman}\ and\ \citenamefont
  {Libermann}(1979)}]{Feldman1979}%
  \BibitemOpen
  \bibfield  {author} {\bibinfo {author} {\bibfnamefont {M.~W.}\ \bibnamefont
  {Feldman}}\ and\ \bibinfo {author} {\bibfnamefont {U.}~\bibnamefont
  {Libermann}},\ }\bibfield  {title} {\bibinfo {title} {On the number of stable
  equilibria and the simultaneous stability of fixation and polymorphism in
  two-locus models},\ }\href {https://doi.org/10.1093/genetics/92.4.1355}
  {\bibfield  {journal} {\bibinfo  {journal} {Genetics}\ }\textbf {\bibinfo
  {volume} {92}},\ \bibinfo {pages} {1355–1360} (\bibinfo {year}
  {1979})}\BibitemShut {NoStop}%
\bibitem [{\citenamefont {Lewontin}\ and\ \citenamefont
  {Kojima}(1960)}]{Lewontin1960}%
  \BibitemOpen
  \bibfield  {author} {\bibinfo {author} {\bibfnamefont {R.~C.}\ \bibnamefont
  {Lewontin}}\ and\ \bibinfo {author} {\bibfnamefont {K.}~\bibnamefont
  {Kojima}},\ }\bibfield  {title} {\bibinfo {title} {The evolutionary dynamics
  of complex polymorphisms},\ }\href {https://doi.org/10.2307/2405995}
  {\bibfield  {journal} {\bibinfo  {journal} {Evolution}\ }\textbf {\bibinfo
  {volume} {14}},\ \bibinfo {pages} {458} (\bibinfo {year} {1960})}\BibitemShut
  {NoStop}%
\bibitem [{\citenamefont {Chakraborty}\ and\ \citenamefont
  {Chakraborty}(2023)}]{Chakraborty2023}%
  \BibitemOpen
  \bibfield  {author} {\bibinfo {author} {\bibfnamefont {S.}~\bibnamefont
  {Chakraborty}}\ and\ \bibinfo {author} {\bibfnamefont {S.}~\bibnamefont
  {Chakraborty}},\ }\bibfield  {title} {\bibinfo {title}
  {Selection-recombination-mutation dynamics: Gradient, limit cycle, and closed
  invariant curve},\ }\href {http://dx.doi.org/10.1103/PhysRevE.108.064404}
  {\bibfield  {journal} {\bibinfo  {journal} {Phys. Rev. E}\ }\textbf {\bibinfo
  {volume} {108}},\ \bibinfo {pages} {064404} (\bibinfo {year}
  {2023})}\BibitemShut {NoStop}%
\bibitem [{\citenamefont {Hastings}(1981)}]{Hastings1981}%
  \BibitemOpen
  \bibfield  {author} {\bibinfo {author} {\bibfnamefont {A.}~\bibnamefont
  {Hastings}},\ }\bibfield  {title} {\bibinfo {title} {Stable cycling in
  discrete-time genetic models.},\ }\href
  {https://doi.org/10.1073/pnas.78.11.7224} {\bibfield  {journal} {\bibinfo
  {journal} {Proc. Natl. Acad. Sci. U.S.A}\ }\textbf {\bibinfo {volume} {78}},\
  \bibinfo {pages} {7224–7225} (\bibinfo {year} {1981})}\BibitemShut
  {NoStop}%
\bibitem [{\citenamefont {Hastings}(1985)}]{Hastings1985}%
  \BibitemOpen
  \bibfield  {author} {\bibinfo {author} {\bibfnamefont {A.}~\bibnamefont
  {Hastings}},\ }\bibfield  {title} {\bibinfo {title} {Four simultaneously
  stable polymorphic equilibria in two-locus two-allele models},\ }\href
  {https://doi.org/10.1093/genetics/109.1.255} {\bibfield  {journal} {\bibinfo
  {journal} {Genetics}\ }\textbf {\bibinfo {volume} {109}},\ \bibinfo {pages}
  {255–261} (\bibinfo {year} {1985})}\BibitemShut {NoStop}%
\bibitem [{\citenamefont {Nagylaki}(1989)}]{Nagylaki1989}%
  \BibitemOpen
  \bibfield  {author} {\bibinfo {author} {\bibfnamefont {T.}~\bibnamefont
  {Nagylaki}},\ }\bibfield  {title} {\bibinfo {title} {The maintenance of
  genetic variability in two-locus models of stabilizing selection.},\ }\href
  {https://doi.org/10.1093/genetics/122.1.235} {\bibfield  {journal} {\bibinfo
  {journal} {Genetics}\ }\textbf {\bibinfo {volume} {122}},\ \bibinfo {pages}
  {235–248} (\bibinfo {year} {1989})}\BibitemShut {NoStop}%
\bibitem [{\citenamefont {Rice}(1961)}]{Rice1961}%
  \BibitemOpen
  \bibfield  {author} {\bibinfo {author} {\bibfnamefont {S.~H.}\ \bibnamefont
  {Rice}},\ }\href@noop {} {\emph {\bibinfo {title} {Evolutionary Theory:
  Mathematical and Conceptual Foundations}}}\ (\bibinfo  {publisher} {Sinauer
  Associates},\ \bibinfo {year} {1961})\BibitemShut {NoStop}%
\bibitem [{\citenamefont {Hofbauer}\ and\ \citenamefont
  {Sigmund}(1998)}]{hofbauer1998}%
  \BibitemOpen
  \bibfield  {author} {\bibinfo {author} {\bibfnamefont {J.}~\bibnamefont
  {Hofbauer}}\ and\ \bibinfo {author} {\bibfnamefont {K.}~\bibnamefont
  {Sigmund}},\ }\href@noop {} {\emph {\bibinfo {title} {Evolutionary games and
  population dynamics}}}\ (\bibinfo  {publisher} {Cambridge university press},\
  \bibinfo {year} {1998})\BibitemShut {NoStop}%
\bibitem [{\citenamefont {Shahshahani}(1979)}]{shahshahani1979new}%
  \BibitemOpen
  \bibfield  {author} {\bibinfo {author} {\bibfnamefont {S.}~\bibnamefont
  {Shahshahani}},\ }\href@noop {} {\emph {\bibinfo {title} {A new mathematical
  framework for the study of linkage and selection}}}\ (\bibinfo  {publisher}
  {American Mathematical Soc.},\ \bibinfo {year} {1979})\BibitemShut {NoStop}%
\bibitem [{\citenamefont {Akin}(2013)}]{akin2013geometry}%
  \BibitemOpen
  \bibfield  {author} {\bibinfo {author} {\bibfnamefont {E.}~\bibnamefont
  {Akin}},\ }\href@noop {} {\emph {\bibinfo {title} {The geometry of population
  genetics}}},\ Vol.~\bibinfo {volume} {31}\ (\bibinfo  {publisher} {Springer
  Science \& Business Media},\ \bibinfo {year} {2013})\BibitemShut {NoStop}%
\bibitem [{\citenamefont {Cressman}(2003)}]{cressman2003book}%
  \BibitemOpen
  \bibfield  {author} {\bibinfo {author} {\bibfnamefont {R.}~\bibnamefont
  {Cressman}},\ }\href@noop {} {\emph {\bibinfo {title} {Evolutionary dynamics
  and extensive form games}}},\ Vol.~\bibinfo {volume} {5}\ (\bibinfo
  {publisher} {MIT Press},\ \bibinfo {year} {2003})\BibitemShut {NoStop}%
\bibitem [{\citenamefont {Neumann}\ and\ \citenamefont
  {Morgenstern}(1944)}]{vonneumann1944}%
  \BibitemOpen
  \bibfield  {author} {\bibinfo {author} {\bibfnamefont {J.}~\bibnamefont
  {Neumann}}\ and\ \bibinfo {author} {\bibfnamefont {O.}~\bibnamefont
  {Morgenstern}},\ }\href {http://www.jstor.org/stable/j.ctt1r2gkx} {\emph
  {\bibinfo {title} {Theory of Games and Economic Behavior}}}\ (\bibinfo
  {publisher} {Princeton University Press},\ \bibinfo {year}
  {1944})\BibitemShut {NoStop}%
\bibitem [{\citenamefont {Hofbauer}\ \emph {et~al.}(1979)\citenamefont
  {Hofbauer}, \citenamefont {Schuster},\ and\ \citenamefont
  {Sigmund}}]{Hofbauer1979}%
  \BibitemOpen
  \bibfield  {author} {\bibinfo {author} {\bibfnamefont {J.}~\bibnamefont
  {Hofbauer}}, \bibinfo {author} {\bibfnamefont {P.}~\bibnamefont {Schuster}},\
  and\ \bibinfo {author} {\bibfnamefont {K.}~\bibnamefont {Sigmund}},\
  }\bibfield  {title} {\bibinfo {title} {A note on evolutionary stable
  strategies and game dynamics},\ }\href
  {https://doi.org/10.1016/0022-5193(79)90058-4} {\bibfield  {journal}
  {\bibinfo  {journal} {J. Theor. Biol.}\ }\textbf {\bibinfo {volume} {81}},\
  \bibinfo {pages} {609–612} (\bibinfo {year} {1979})}\BibitemShut {NoStop}%
\bibitem [{\citenamefont {Harsanyi}(1968{\natexlab{a}})}]{Harsanyi1968part2}%
  \BibitemOpen
  \bibfield  {author} {\bibinfo {author} {\bibfnamefont {J.~C.}\ \bibnamefont
  {Harsanyi}},\ }\bibfield  {title} {\bibinfo {title} {Games with incomplete
  information played by “bayesian” players, part ii. bayesian equilibrium
  points},\ }\href {https://doi.org/10.1287/mnsc.14.5.320} {\bibfield
  {journal} {\bibinfo  {journal} {Manag. Sci.}\ }\textbf {\bibinfo {volume}
  {14}},\ \bibinfo {pages} {320–334} (\bibinfo {year}
  {1968}{\natexlab{a}})}\BibitemShut {NoStop}%
\bibitem [{\citenamefont {Cover}\ and\ \citenamefont
  {Thomas}(2005)}]{Cover2005}%
  \BibitemOpen
  \bibfield  {author} {\bibinfo {author} {\bibfnamefont {T.~M.}\ \bibnamefont
  {Cover}}\ and\ \bibinfo {author} {\bibfnamefont {J.~A.}\ \bibnamefont
  {Thomas}},\ }\href {https://doi.org/10.1002/047174882x} {\emph {\bibinfo
  {title} {Elements of Information Theory}}}\ (\bibinfo  {publisher} {Wiley},\
  \bibinfo {year} {2005})\BibitemShut {NoStop}%
\bibitem [{\citenamefont {Wiggins}(2003)}]{Wiggins2003}%
  \BibitemOpen
  \bibfield  {author} {\bibinfo {author} {\bibfnamefont {S.}~\bibnamefont
  {Wiggins}},\ }\href@noop {} {\emph {\bibinfo {title} {Introduction to Applied
  Nonlinear Dynamical Systems and Chaos}}}\ (\bibinfo  {publisher}
  {Springer-Verlag},\ \bibinfo {year} {2003})\BibitemShut {NoStop}%
\bibitem [{\citenamefont {Tilman}\ \emph {et~al.}(2023)\citenamefont {Tilman},
  \citenamefont {Vasconcelos}, \citenamefont {Ak\c{c}ay},\ and\ \citenamefont
  {Plotkin}}]{Tilman2023}%
  \BibitemOpen
  \bibfield  {author} {\bibinfo {author} {\bibfnamefont {A.~R.}\ \bibnamefont
  {Tilman}}, \bibinfo {author} {\bibfnamefont {V.~V.}\ \bibnamefont
  {Vasconcelos}}, \bibinfo {author} {\bibfnamefont {E.}~\bibnamefont
  {Ak\c{c}ay}},\ and\ \bibinfo {author} {\bibfnamefont {J.~B.}\ \bibnamefont
  {Plotkin}},\ }\bibfield  {title} {\bibinfo {title} {The evolution of
  forecasting for decision-making in dynamic environments},\ }\href
  {http://dx.doi.org/10.1177/26339137231221726} {\bibfield  {journal} {\bibinfo
   {journal} {Collective Intelligence}\ }\textbf {\bibinfo {volume} {2}},\
  \bibinfo {pages} {1–14} (\bibinfo {year} {2023})}\BibitemShut {NoStop}%
\bibitem [{\citenamefont {Rand}\ \emph {et~al.}(2017)\citenamefont {Rand},
  \citenamefont {Tomlin}, \citenamefont {Bear}, \citenamefont {Ludvig},\ and\
  \citenamefont {Cohen}}]{Rand2017}%
  \BibitemOpen
  \bibfield  {author} {\bibinfo {author} {\bibfnamefont {D.~G.}\ \bibnamefont
  {Rand}}, \bibinfo {author} {\bibfnamefont {D.}~\bibnamefont {Tomlin}},
  \bibinfo {author} {\bibfnamefont {A.}~\bibnamefont {Bear}}, \bibinfo {author}
  {\bibfnamefont {E.~A.}\ \bibnamefont {Ludvig}},\ and\ \bibinfo {author}
  {\bibfnamefont {J.~D.}\ \bibnamefont {Cohen}},\ }\bibfield  {title} {\bibinfo
  {title} {Cyclical population dynamics of automatic versus controlled
  processing: An evolutionary pendulum.},\ }\href
  {https://doi.org/10.1037/rev0000079} {\bibfield  {journal} {\bibinfo
  {journal} {Psychol. Rev.}\ }\textbf {\bibinfo {volume} {124}},\ \bibinfo
  {pages} {626–642} (\bibinfo {year} {2017})}\BibitemShut {NoStop}%
\bibitem [{\citenamefont {Barfar}(2022)}]{Barfar2022}%
  \BibitemOpen
  \bibfield  {author} {\bibinfo {author} {\bibfnamefont {A.}~\bibnamefont
  {Barfar}},\ }\bibfield  {title} {\bibinfo {title} {A
  linguistic/game-theoretic approach to detection/explanation of propaganda},\
  }\href {https://doi.org/10.1016/j.eswa.2021.116069} {\bibfield  {journal}
  {\bibinfo  {journal} {Expert Syst. Appl.}\ }\textbf {\bibinfo {volume}
  {189}},\ \bibinfo {pages} {116069} (\bibinfo {year} {2022})}\BibitemShut
  {NoStop}%
\bibitem [{\citenamefont {Nowak}\ \emph {et~al.}(2004)\citenamefont {Nowak},
  \citenamefont {Sasaki}, \citenamefont {Taylor},\ and\ \citenamefont
  {Fudenberg}}]{Nowak2004}%
  \BibitemOpen
  \bibfield  {author} {\bibinfo {author} {\bibfnamefont {M.~A.}\ \bibnamefont
  {Nowak}}, \bibinfo {author} {\bibfnamefont {A.}~\bibnamefont {Sasaki}},
  \bibinfo {author} {\bibfnamefont {C.}~\bibnamefont {Taylor}},\ and\ \bibinfo
  {author} {\bibfnamefont {D.}~\bibnamefont {Fudenberg}},\ }\bibfield  {title}
  {\bibinfo {title} {Emergence of cooperation and evolutionary stability in
  finite populations},\ }\href {https://doi.org/10.1038/nature02414} {\bibfield
   {journal} {\bibinfo  {journal} {Nature}\ }\textbf {\bibinfo {volume}
  {428}},\ \bibinfo {pages} {646–650} (\bibinfo {year} {2004})}\BibitemShut
  {NoStop}%
\bibitem [{\citenamefont {Ohtsuki}\ and\ \citenamefont
  {Nowak}(2006)}]{Ohtsuki2006}%
  \BibitemOpen
  \bibfield  {author} {\bibinfo {author} {\bibfnamefont {H.}~\bibnamefont
  {Ohtsuki}}\ and\ \bibinfo {author} {\bibfnamefont {M.~A.}\ \bibnamefont
  {Nowak}},\ }\bibfield  {title} {\bibinfo {title} {The replicator equation on
  graphs},\ }\href {https://doi.org/10.1016/j.jtbi.2006.06.004} {\bibfield
  {journal} {\bibinfo  {journal} {J. Theor. Biol.}\ }\textbf {\bibinfo {volume}
  {243}},\ \bibinfo {pages} {86} (\bibinfo {year} {2006})}\BibitemShut
  {NoStop}%
\bibitem [{\citenamefont {Harsanyi}(1967)}]{Harsanyi1967part1}%
  \BibitemOpen
  \bibfield  {author} {\bibinfo {author} {\bibfnamefont {J.~C.}\ \bibnamefont
  {Harsanyi}},\ }\bibfield  {title} {\bibinfo {title} {Games with incomplete
  information played by “bayesian” players, part i. the basic model},\
  }\href {https://doi.org/10.1287/mnsc.14.3.159} {\bibfield  {journal}
  {\bibinfo  {journal} {Manag. Sci.}\ }\textbf {\bibinfo {volume} {14}},\
  \bibinfo {pages} {159–182} (\bibinfo {year} {1967})}\BibitemShut {NoStop}%
\bibitem [{\citenamefont {Harsanyi}(1968{\natexlab{b}})}]{Harsanyi1968part3}%
  \BibitemOpen
  \bibfield  {author} {\bibinfo {author} {\bibfnamefont {J.~C.}\ \bibnamefont
  {Harsanyi}},\ }\bibfield  {title} {\bibinfo {title} {Games with incomplete
  information played by "bayesian" players, part iii. the basic probability
  distribution of the game},\ }\href {https://doi.org/10.1287/mnsc.14.7.486}
  {\bibfield  {journal} {\bibinfo  {journal} {Manag. Sci.}\ }\textbf {\bibinfo
  {volume} {14}},\ \bibinfo {pages} {486} (\bibinfo {year}
  {1968}{\natexlab{b}})}\BibitemShut {NoStop}%
\end{thebibliography}%
\end{document}